\documentclass[lettersize,journal]{IEEEtran}
\usepackage{amsmath,amsfonts}
\usepackage{algorithmic}
\usepackage{algorithm}
\usepackage{array}
\usepackage[caption=false,font=normalsize,labelfont=sf,textfont=sf]{subfig}
\usepackage{textcomp}
\usepackage{stfloats}
\usepackage{url}
\usepackage{verbatim}
\usepackage{graphicx}
\usepackage{cite}
\usepackage{xcolor}
\usepackage{enumitem}

\usepackage{hyperref}

\newcommand{\ignore}[1]{}

\newcommand{\zztitle}[1]{\vspace{2pt}\noindent\textbf{#1.}}

\newcommand{\apiname}[1]{\textit{\texttt{#1}}}

\newcommand{\mytcp}{\texttt{PnO-TCP}}
\newcommand{\mytcph}{\texttt{TCP-Proxy}}
\newcommand{\mytcpn}{\texttt{TCP-Bridge}}
\newcommand{\myname}{\texttt{PnO}}
\newcommand{\mynamep}{\texttt{PnO-Shim}}

\begin{document}

\title{Plug \& Offload: Transparently Offloading TCP Stack onto Off-path SmartNIC with \mytcp}

\author{
{\rm Hailong Nan, Zhou Zhe, Min Yang}\\
Department of Computer Science, Fudan University
} 



\maketitle

\begin{abstract}



Host CPU resources are heavily consumed by TCP stack processing, limiting scalability in data centers. Existing offload methods typically address only partial functionality or lack flexibility.


This paper introduces \myname{} (Plug \& Offload), an approach to fully offload TCP processing transparently onto off-path SmartNICs (NVIDIA BlueField DPUs). Key to our solution is \mytcp{}, a novel TCP stack specifically designed for efficient execution on the DPU's general-purpose cores, spanning both the host and the SmartNIC to facilitate the offload. \mytcp{} leverages a lightweight, user-space stack based on DPDK, achieving high performance despite the relatively modest computational power of off-path SmartNIC cores.

Our evaluation, using real-world applications (Redis, Lighttpd, and HAProxy), demonstrates that \myname{} achieves transparent TCP stack offloading, leading to both substantial reductions in host CPU usage and, in many cases, significant performance improvements, particularly for small packet scenarios (\textless~2KB) where RPS gains of 34\%-127\% were observed in single-threaded tests.



\end{abstract}


\section{Introduction}

\IEEEPARstart{M}{odern} data centers face escalating demands driven by dramatically increasing network data rates and evolving applications~\cite{firestone2018azure}. This surge places a significant burden on host CPUs, as protocol stack processing consumes a non-negligible amount of expensive compute cycles. To alleviate this bottleneck and enhance overall processing speed and energy efficiency, SmartNICs~\cite{agilio_smartnic,NVIDIA_BlueField2_Datasheet,what-is-a-smartnic} have emerged as critical components, designed to take responsibility for packet processing. Particularly, off-path SmartNICs (often termed DPUs~\cite{NVIDIA_BlueField2_Datasheet, NVIDIA_BlueField3_Datasheet,Broadcom_P225P}) combine specialized hardware accelerators with multiple general-purpose cores. While this architecture offers improved energy efficiency, the generally lower per-core performance of these embedded cores compared to host CPUs presents both an opportunity and a challenge for offloading complex tasks.

SmartNICs offer a compelling solution to reduce host CPU overhead by offloading tasks, conserving valuable compute cycles. Indeed, successful offloads have been demonstrated for diverse functions like Network Functions (NF)\cite{lin2020panic}, Key-Value stores (KV)\cite{li2017kv}, and Intrusion Detection Systems (IDS)\cite{kaufmann2016high}, showcasing the potential of these devices. However, when applied to the crucial task of TCP processing, existing offloading strategies often fall short. They typically address only narrow aspects – either specific scenarios like accelerating connection setup/teardown (e.g., AccelTCP\cite{moon2020acceltcp}) or basic, fixed functions such as checksum calculation and TCP Segmentation Offload (TSO)\cite{shinde2013we}. This limited scope restricts both their flexibility for protocol evolution and their general applicability across diverse workloads.

Seeking a more comprehensive solution beyond partial offloads, TCP Offload Engines (TOEs)\cite{chiang2010full,wu2006design} represented an ambitious attempt to move the entire TCP/IP stack onto dedicated NIC hardware. However, this ambition largely failed to translate into widespread adoption, primarily because the fixed nature of these hardware offloads\cite{linuxtoe} severely constrained protocol evolution after deployment~\cite{Danielnsdi18, Martysosp19snap, mogul2003tcp}. More recent efforts leverage programmable NICs, but still face hurdles. While Tonic~\cite{arashloo2020enabling} provides flexible building blocks for FPGA-based SmartNICs, the inherent difficulty and lengthy development cycles of FPGAs limit its practical use. Another approach, FlexTOE~\cite{Rajath2019FlexTOE}, achieves transparent deployment on on-path SmartNICs using Flow Processing Cores (FPCs). Unfortunately, these FPCs often lack crucial computational capabilities (like timers, floating-point, division) and possess weak overall computing power, rendering them unsuitable for control-intensive TCP functions (e.g., calculating an ECN gradient takes 1.9 µs per RTT~\cite{Rajath2019FlexTOE}). Consequently, even FlexTOE ends up offloading only data-path transmissions, leaving core TCP functionalities, connection management, congestion control, retransmission, and timeouts, on the host due to these FPC limitations.

To address the shortcomings of previous partial and inflexible offload methods, this paper introduces Plug \& Offload (\myname{})\footnote{The name is inspired by \textit{Plug and Play} (PnP)}, a novel approach achieving transparent and entire TCP protocol stack offloading onto off-path SmartNICs, specifically NVIDIA BlueField DPUs\footnote{This paper will use "DPU" to specifically refer to off-path SmartNICs.}. Unlike prior attempts constrained by hardware rigidity or weak processing cores, \myname{} leverages the capable general-purpose cores found on these DPUs. It operates by dynamically intercepting standard socket API calls within host applications and seamlessly redirecting them to \mytcp{}, our custom-designed, high-performance TCP stack executing efficiently on the DPU subsystem, thereby realizing full TCP offload without requiring application modifications.



Central to the \myname{} approach is \mytcp{}, a novel TCP stack specifically architected for this cooperative host-DPU execution model. Departing from the standard kernel stack, \mytcp{} is a custom user-space implementation meticulously designed to address the unique challenges of DPU offloading. Its core principles involve: (1) Efficiency on DPU Cores: It is engineered for high-performance and low-latency execution specifically on the DPU's ARM cores, leveraging parallelism across multiple cores. (2) Minimized Host-DPU Communication: \mytcp{} incorporates a custom, optimized communication mechanism utilizing shared memory and carefully managed DMA operations. This minimizes the overhead associated with transferring data and synchronizing state across the PCIe bus between the host and the DPU, which is critical for overall performance.


Despite the promise of \myname{} and the design principles of \mytcp{}, achieving transparent and efficient full TCP stack offloading in this manner presents significant hurdles. Three key challenges must be overcome: (1) Weaker DPU Core Performance: The general-purpose cores on off-path SmartNICs, while numerous and power-efficient, typically offer lower single-core performance compared to host CPUs, demanding a highly optimized stack. (2) High PCIe Latency: Communication between the host and the SmartNIC over the PCIe bus introduces substantial latency, impacting responsiveness. (3) Costly Host-Initiated Operations: TCP operations initiated by the host application can incur significant delays due to the multiple DMA transfers required for data and control exchange. \mytcp{} is specifically engineered to tackle these challenges directly through its lightweight and optimized stack design, the use of carefully structured communication queue rings to manage data flow and minimize DMA overhead, and leveraging the Data Plane Development Kit (DPDK) for highly efficient low-level packet processing on the DPU subsystem.


Our evaluation, using real-world applications including Redis, Lighttpd, and HAProxy, demonstrates that \myname{} successfully achieves transparent TCP stack offloading. This results in substantial reductions in host CPU usage and, in many cases, significant performance improvements, particularly for small-packet scenarios (\textless 2KB) where gains of 34\%-127\% were observed in single-threaded tests. \myname{} unlocks the potential of SmartNICs for a wider range of applications by removing the barriers to adoption, paving the way for more efficient and scalable data center infrastructure.

The main contributions of this paper are as follows:
\begin{itemize}[leftmargin=*]
\item We propose \myname{}, a first transparent offloading mechanism that enables the entire TCP stack to be offloaded to the DPU's off-path without requiring any modifications or recompilation of existing applications.

\item We design \mytcp{}, a TCP stack spanning both the host and the DPU, specifically optimized for offloading the host TCP stack.

\item We evaluate \myname{} with real-world applications, including Redis, HAProxy, and Lighttpd, demonstrating its feasibility, correctness, and host CPU savings.
\end{itemize}

\section{Background and Motivation}
\subsection{Background}

\begin{figure}[ht]
    \centering
    \includegraphics[width=0.45\textwidth]{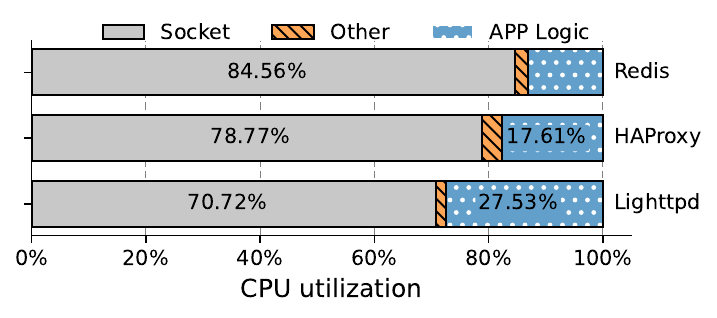}
    \caption{CPU Utilization Breakdown of Real-World Network Applications.}
    \label{fig:redis_flame}
\end{figure}

\begin{figure}[ht]
    \centering
    \includegraphics[width=0.40\textwidth]{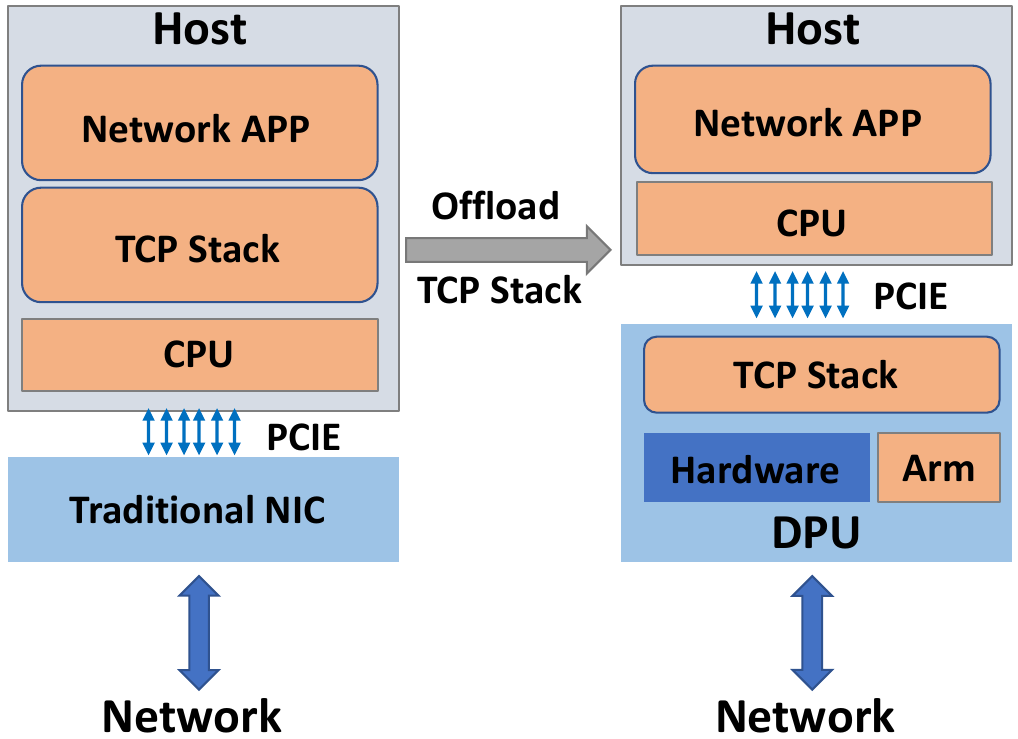}
    \caption{Comparison of Traditional Host TCP Stack vs. Offloaded TCP Stack on DPU.}
    \label{fig:offload_tcp_stack}
\end{figure}

\textbf{The TCP stack running on the CPU exhibits low efficiency and high consumption of general-purpose computing resources.} 
The Transmission Control Protocol (TCP)~\cite{rfc793}, a cornerstone of modern networking, ensures reliable and ordered data delivery between applications.Yet, its implementation in the TCP stack imposes substantial computational overhead, heavily taxing host CPU resources to manage network traffic, as depicted in the left part of Fig.~\ref{fig:offload_tcp_stack}. As network speeds surge~\cite{firestone2018azure}, this stack frequently emerges as a bottleneck for high-throughput applications. Fig.~\ref{fig:redis_flame} reveals that application-level processing typically consumes only a minor portion of CPU time, with over 70\% on average devoured by TCP stack and socket operations.

\begin{figure}

    \centering
    \includegraphics[width=0.8\columnwidth]{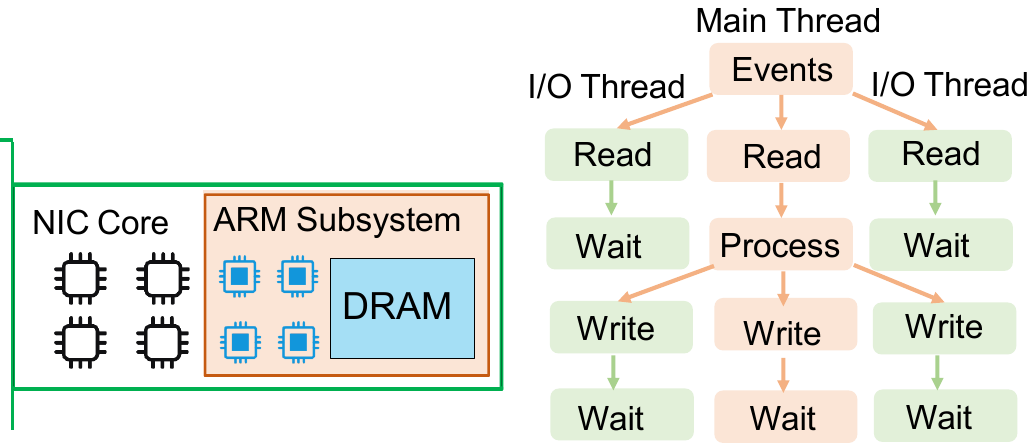}
    \caption{On the left is the SmartNIC depicted, and on the right is the Redis Thread Model.}
    \label{fig:redis_model}

\end{figure}

To mitigate this overhead and improve application performance, developers often employ multi-threaded networking I/O. This approach leverages parallelism by distributing the network workload across multiple CPU cores. For example, Redis has adopted a multi-threaded model specifically for I/O operations.  As illustrated in the Fig.~\ref{fig:redis_model} (right side), Redis uses dedicated networking threads that act as consumers, exclusively handling I/O tasks, while the main thread focuses on key-value retrieval and acts as a producer. 
When an I/O event is detected, the main thread retrieves and dispatches it to different I/O threads to deal.

\textbf{Offloading the network stack to conserve valuable host CPU resources is a compelling approach, increasingly enabled by off-path SmartNICs, particularly Data Processing Units (DPUs)\cite{Broadcom_P225P, NVIDIA_BlueField3_Datasheet, NVIDIA_BlueField2_Datasheet}.} DPUs are rapidly gaining prominence as a significant market segment\cite{dpu_smartnic_market}, driven by their ability to address critical I/O performance bottlenecks in the post-Moore's Law era~\cite{10.1145/3282307}. Architecturally, DPUs integrate a programmable multi-core System-on-Chip (SoC) with DRAM, positioned adjacent to the primary NIC cores but operating outside the critical data path (Fig.\ref{fig:redis_model} left). Using the Nvidia BlueField DPU as a representative example (using "DPU" and "off-path SmartNIC" interchangeably in this paper), their key advantage lies in offloading infrastructure services like networking, storage, and security. This is achieved through a powerful combination: First, specialized hardware accelerators boost performance for critical functions, reducing host overhead and improving I/O throughput. Second, and crucially for flexibility, DPUs feature general-purpose programmable cores. This programmability allows for custom processing logic directly on the NIC, distinguishing DPUs from fixed-function offloads, and is essential for executing complex, stateful tasks such as the entire TCP stack. As illustrated (Fig.\ref{fig:offload_tcp_stack} right), running the complete TCP stack on these DPU cores can substantially reduce host CPU consumption and potentially increase throughput, especially for small packets. Therefore, leveraging this unique synergy of hardware acceleration and programmability within DPUs offers a promising path towards enhanced system efficiency and performance.

\textbf{Although DPUs are fundamentally capable of network function offloading, the specific exploration of fully offloading the TCP stack onto their general-purpose cores remains limited.} Current research predominantly leverages DPUs' integrated hardware accelerators or targets only partial stack functions (e.g., TLS acceleration, compression~\cite{Pismenny_Eran_Yehezkel_Liss_Morrison_Tsafrir_2021,Li_Kashyap_Guo_Lu_2024}), rather than utilizing the programmable cores for the complete TCP protocol. Yet, this overlooks the significant potential of these cores. DPUs feature energy-efficient, general-purpose ARM cores (like the 16 Cortex-A78s in BlueField-3~\cite{NVIDIA_BlueField3_Datasheet}) offering superior power efficiency and high density (e.g., up to 36 ARM/48 MIPS cores in OCTEON\cite{Marvell_OCTEON_10_Datasheet}). While having lower per-core IPC than host x86, their aggregate performance and efficiency make them well-suited for networking workloads. Consequently, leveraging these programmable DPU cores for complete TCP stack processing is a highly relevant but underexplored avenue for enhancing network efficiency. Furthermore, existing partial offloading approaches~\cite{Suresh_Michalowicz_Ramesh_Contini_Yao_Xu_Shafi_Subramoni_Panda_2023,moon2020acceltcp,Kim_Ng_Gong_Kwon_Yu_Park_2023} often lack generality and demand significant, DPU-specific application restructuring, highlighting the need for a more comprehensive and transparent solution.

\subsection{ Related Work }

Extensive research has been conducted on network function offloading, with particular attention to offloading TCP-related tasks and other network functions to NICs. This section reviews key approaches in TCP stack offloading and SmartNIC-based offloading techniques.

\zztitle{TCP stack offload}
Several studies have explored accelerating specific components of the TCP stack without fully offloading it. Partial TCP offload techniques typically focus on fixed-function accelerations, such as TCP Segmentation Offload (TSO), TCP/IP checksum offload, and Large Receive Offload (LRO).  More advanced approaches, such as AccelTCP~\cite{moon2020acceltcp}, propose offloading TCP connection setup and teardown to NICs, effectively reducing computational overhead for short-lived connections. However, AccelTCP relies on on-path SmartNICs, keeping the majority of the TCP stack processing on the host CPU. Other works have investigated offloading large message transfers to NICs~\cite{kim2006connection}, but these solutions remain limited in scope. TCP Offload Engines (TOEs)\cite{chiang2010full, wu2006design} represent a more comprehensive approach by attempting to offload the entire TCP/IP stack onto dedicated NIC hardware\cite{arashloo2020enabling, pratt2001arsenic, putnam2014reconfigurable}. Tonic~\cite{arashloo2020enabling} demonstrates in simulation that flexible, high-performance TCP transmission offload is possible. However, it does not implement a full TCP data-path offload (including receiver-side processing) in a real-world deployment. Despite their potential benefits, TOEs have not been widely adopted due to their tight coupling with hardware implementations, invasive modifications to the kernel stack, and the limited compute resources available on NICs~\cite{LWN_Corbet_Kernel_Module_Loading}. Additionally, TOEs suffer from restricted operational flexibility, as firmware updates are required to fix bugs, update congestion control algorithms, or introduce new TCP options.

\zztitle{SmartNIC offload}
Research has also extensively explored offloading various computationally intensive network functions, distinct from the TCP stack itself, onto SmartNICs, often leveraging specialized hardware or targeting specific application domains.
For instance, cryptographic offloads are common, with kTLS~\cite{pismenny2016tls}, SmartTLS~\cite{kim2020case}, and IPsec offload~\cite{agrawal2012performance} primarily utilizing dedicated accelerators, though sometimes requiring application changes. Packet processing and forwarding have also seen offloads, using GPUs for routing lookups~\cite{han2010packetshader, li2013gamt, vasiliadis2014gaspp} or employing NIC-level processing with eBPF/XDP~\cite{wang2023oxdp}.
In the storage domain, SmartNICs accelerate operations by offloading file system logic (LineFS~\cite{kim2021linefs}), multi-tenant storage functions (Hyperloop~\cite{kim2018hyperloop}), or key-value store computations~\cite{zhang2020fpga, li2017kv}. Even approaches closer to I/O handling, like IO-TCP~\cite{Kim_Ng_Gong_Kwon_Yu_Park_2023}, focus narrowly on specific patterns (disk writes), using accelerators and DPDK for data plane tasks while keeping control logic on the host.
Crucially, the vast majority of these function-specific offloads either rely primarily on hardware accelerators, target only parts of the network pipeline, lack generality, or require application modifications.

\subsection{Motivation}

Existing TCP offload techniques face critical limitations. Whether relying on fixed-function accelerators or constrained on-path cores (like FPCs lacking essential features like timers), they typically handle only partial stack functionality and lack the flexibility needed for protocol evolution. Rigid hardware offloads (TOEs) stifle innovation by requiring firmware updates for changes, while complex partitioning for partial offloads restricts scalability. This gap highlights the need for a more comprehensive solution. The rise of off-path DPUs, equipped with versatile programmable ARM cores, presents a significant opportunity. Capable of executing full software stacks, these DPUs enable complete TCP offloading, overcoming the inflexibility and partial nature of prior methods. This promises both reduced host CPU overhead and potential performance boosts. Therefore, the primary motivation for this work is to develop a transparent and efficient approach that fully leverages this DPU potential, enabling seamless, complete TCP offload without requiring application modification.


\section{Goals, Challenges, and Opportunities}


Prior research has explored various approaches to offloading TCP processing to SmartNICs, but these methods often suffer from limited transparency and incomplete offloading, hindering widespread adoption. To this end, this paper focuses on achieving practical, transparent, and efficient full TCP stack offloading to DPUs. This section first outlines the design goals for our full TCP stack offloading mechanism. We then identify and discuss the key challenges and limitations of executing the entire TCP stack on a DPU. Finally, we explore potential opportunities to mitigate these limitations and enhance the effectiveness of our proposed solution.

\subsection{Goals}
Our objective is to explore the feasibility of offloading the entire TCP stack to a DPU, thereby reducing the computational burden on the host CPU. To achieve this, we propose the following design goals:
\begin{itemize}[leftmargin=*]

\item  \textbf{Generalized Full TCP Offloading:} Our approach offloads the entire TCP stack to the DPU in a generalized manner, avoiding application-specific or scenario-specific solutions.

\item  \textbf{Transparency to Developers:} The offloading mechanism is fully transparent, requiring no application modifications or manual refactoring by developers.

\item  \textbf{Efficient Host CPU Utilization and DPU Resources:} Our primary goal is to reduce host CPU utilization while leveraging DPU computational resources effectively, achieving both CPU savings and performance improvements despite the DPU ARM cores’ limitations.
\end{itemize}

To realize the aforementioned objectives, we propose a novel TCP stack architecture that spans both the host and DPU. This implementation must satisfy the following key requirements: \textcircled{1} seamless integration between the host and DPU, facilitating efficient TCP stack offloading, \textcircled{2} minimal additional CPU overhead on the host compared to a non-offloaded setup, \textcircled{3} high performance, achieving higher throughput than the Linux TCP stack on the DPU’s ARM subsystem, and \textcircled{4} high scalability, allowing flexible adaptation to different core configurations.

\subsection{Challenges} \label{sec:challenges}

To achieve complete TCP offload, the full TCP protocol stack must execute on the DPU's general-purpose cores. This fundamental requirement introduces several critical challenges:

\zztitle{Challenge 1: Computational Limitations of DPU Cores for Full TCP Stack Execution.} A key obstacle in offloading the TCP stack to the DPU lies in the limited computational capacity of its general-purpose cores, such as the ARM A78 used in our NVIDIA BlueField DPU. The standard Linux kernel TCP stack, known for being resource-intensive, is ill-suited for the DPU's energy-efficient but computationally less powerful cores, leading to inefficient execution and performance bottlenecks. For instance, Fig.~\ref{fig:redis_flame} illustrates that TCP processing consumes over 84.56\% of the workload in Redis on a host CPU, highlighting the stack’s computational demands. Our experiments further reveal that the x86 AMD EPYC 7302 host core delivers more 2× higher single-core performance than the ARM A78 DPU core. This substantial performance disparity complicates the goal of running the full TCP stack on the DPU while maintaining throughput and latency comparable to host-based execution, posing a significant challenge to effective offloading.

\zztitle{Challenge 2: Communication Overhead Between the Host and DPU} TCP stack offloading necessitates frequent communication between the host and the DPU, as the DPU must handle TCP services on behalf of the host. As shown in the Fig.~\ref{fig:offload_tcp_stack}, this communication primarily relies on DMA transfers over the PCIe bus. However, PCIe introduces considerable latency; our measurements indicate that a single DMA operation takes approximately 2.1 µs, limiting throughput to around 430K transactions per second. This level of latency is unacceptable for high-performance networking applications, making it crucial to explore optimizations that reduce communication overhead.

\zztitle{Challenge 3: End-to-End Latency of TCP Operations} The end-to-end latency of TCP operations presents another significant challenge. When a TCP operation is initiated on the host, data must first be transferred via DMA to the DPU subsystem, where it is processed by the TCP stack before being written to the protocol stack's backend buffer. The result must then be sent back to the host. This sequence of operations incurs substantial processing time, approximately 4.2 µs, largely dominated by the two required DMA transfers (each ~2.1 µs), plus additional memory copy and processing overhead. Such latency is prohibitive for low-latency applications, necessitating further optimizations to improve responsiveness.

\subsection{Opportunities}

The challenges discussed earlier have led many existing studies~\cite{Li_Chen_Shen_Wang_Cao_2025, wei2023characterizing, DBLP:journals/corr/abs-2105-06619, Thostrup2022ADE} to discourage the offloading of an entire TCP stack to the DPU's general-purpose core. However, these challenges also present opportunities for optimization. To address Challenge 1, we draw inspiration from lightweight user-space TCP stacks, such as mTCP~\cite{jeong2014mtcp}, and investigate the feasibility of developing a high-performance TCP stack optimized for DPUs. To tackle Challenges 2 and 3, we explore the characteristics of DOCA DMA~\cite{doca_dma} transfers to minimize and distribute DMA transmission latency effectively.

\zztitle{Userspace TCP Stack Based on DPDK} The high computational overhead of the Linux kernel TCP stack motivates us to explore lightweight alternatives. Implementing a high-performance TCP stack on ARM architectures using DPDK (Data Plane Development Kit)~\cite{dpdk} offers several advantages. DPDK enables direct packet processing in user space, bypassing the kernel networking stack, which significantly reduces latency and improves throughput—especially beneficial for ARM-based DPUs. For example, with a 64B packet size, DPDK-based mTCP achieves about four times the throughput of Linux Kernel TCP~\cite{jeong2014mtcp}.

While multiple DPDK-based TCP stacks exist, including F-Stack~\cite{fstack}, Seastar and mTCP~\cite{jeong2014mtcp}, many of them are highly optimized for multi-core x86 architectures and are not well-suited for ARM-based environments~\cite{F_stack_issue}. However, this presents an opportunity to redesign a high-performance, ARM-optimized TCP stack leveraging DPDK, potentially overcoming the inefficiencies of existing solutions and making TCP stack offloading on DPUs more viable.

\begin{figure}[h!]
    \centering
    \includegraphics[width=0.45\textwidth]{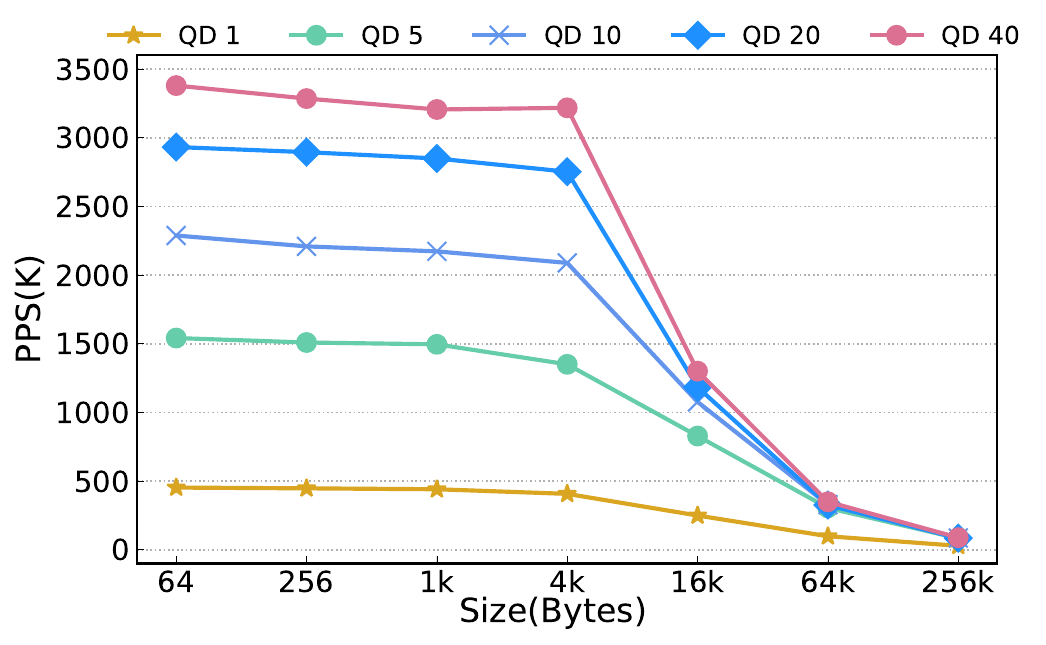}
    \caption{DOCA DMA Performance Analysis Under Different Queue Depths (QD).}
    \label{fig:DOCA}
\end{figure}

\zztitle{Batching DMA Requests to Amortize Latency} Although DMA operations inherently introduce high latency, batching multiple DMA requests into a single operation can significantly reduce the per-request delay. To quantify this effect, we analyzed DOCA DMA~\cite{doca_dma} performance under varying batch sizes and request sizes. As illustrated in the Fig.~\ref{fig:DOCA}, batching allows multiple requests to be processed within a single DMA transaction, leading to notable latency improvements.

For instance, when ten requests are batched together, the amortized average latency per request is reduced to approximately 0.4 µs for a 4K memory size. This finding suggests that issuing multiple DMA requests concurrently can greatly improve performance. Furthermore, as observed in the Fig.~\ref{fig:DOCA}, the requests per second (RPS) remains nearly constant for small request sizes (i.e., less than 4KB). This insight enables us to group multiple small requests into a single large DMA transaction without saturating the available bandwidth, further mitigating the impact of DMA-induced latency.

\section{Design and Implementation}

In this section, we present the design of \myname{} in detail. \myname{} comprises two key components: \mytcp{}, a protocol stack spanning both the host and DPU, and the \mynamep{}, which dynamically offloads the host TCP stack by redirecting TCP socket API. We first provide an overview of \myname{}, followed by an in-depth discussion of the \mynamep{}.

\subsection{\myname{} Overview}
\label{sec:PnO_overview}
\begin{figure}[ht]
    \centering
    \includegraphics[width=0.20\textwidth]{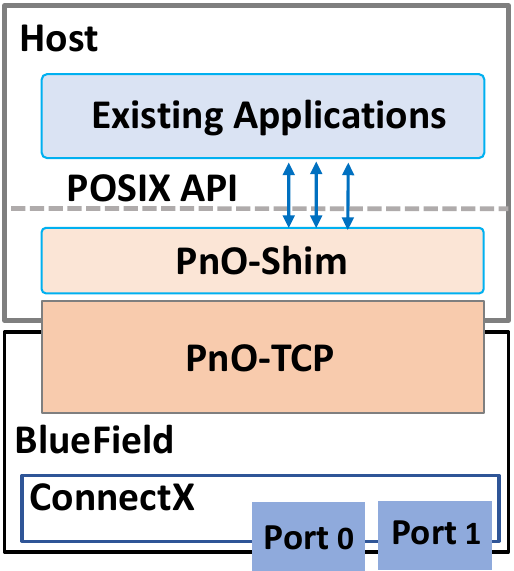}
    \caption{\myname{} Architecture Overview: Transparent TCP Stack Offloading.}
    \label{fig:pno_st}
\end{figure}

The \myname{} architecture, illustrated in the Fig.~\ref{fig:pno_st}, comprises two main components. The \mynamep{} dynamically redirects application network APIs to offloaded TCP APIs, ensuring transparent TCP processing for network applications without requiring modifications. The \mytcp{} stack, a lightweight TCP stack implementation designed for offloading, operates across both the host and DPU. The core protocol stack of \mytcp{} is executed on the DPU, where the actual TCP processing takes place. The integration of the \mynamep{} with \mytcp{} achieves full TCP stack offloading in a manner that is completely transparent to developers.

The data flow operates as follows: when an unmodified network application requiring TCP offloading is executed, the \mynamep{} intercepts all network API calls made by the application. These calls are then forwarded to the \mytcp{} stack for processing. Upon completing the network operations on the DPU, the \mytcp{} stack returns the results to the application running on the host. The following sections detail the design and implementation of the \mynamep{}, while Section~\ref{sec:htcp_overview} elaborates on the \mytcp{} stack.

\subsection{\mynamep{}}

The \mynamep{} facilitates the translation of network-related API calls from applications into corresponding \mytcp{} APIs. Several challenges necessitate the use of the \mynamep{}. Replacing network-related APIs without recompiling the application is essential, but this cannot be achieved simply by substituting shared libraries because: \textcircled{1} Some developers invoke network services directly via system calls rather than shared libraries; \textcircled{2} Certain network-related APIs, such as \apiname{read} and \apiname{write}, are also used for other system services; and \textcircled{3} Network services interact with other host system services through APIs like \apiname{epoll\_wait} (which may monitor both network and other system events) and \apiname{sendfile} (which transfers data from the local file buffer to the network socket stream). To address these challenges, the \mynamep{} enables transparent TCP offloading without requiring modifications to existing applications.

To achieve comprehensive interception, the \mynamep{} captures all potential system calls and library API invocations related to network operations. System calls are intercepted using \apiname{ptrace}, whereas library API calls are intercepted by modifying the Global Offset Table (GOT) of executable files at runtime. For APIs such as \apiname{read} and \apiname{write}, which serve both network and local host services, the \mynamep{} differentiates them based on file descriptors: \mytcp{} allocates file descriptors starting from a high range (e.g., 1000). When the \mynamep{} intercepts an operation on a file descriptor below this range, it classifies it as belonging to local host services, whereas those in the allocated high range are handled via \mytcp{} APIs.

For functions like \apiname{epoll\_wait}, which can monitor both network and local host events, the \mynamep{} employs a non-blocking polling mechanism to separately monitor local events. When performing \apiname{epoll\_ctl} operations, if an event to be monitored belongs to a local host event, the \mynamep{} creates a separate \apiname{epoll} instance to handle local events. Each time \apiname{epoll\_wait} is called, the \mynamep{} retrieves local host events separately, merges these results with network events, and then returns them to the application—ensuring this process remains transparent to the application. Similarly, for operations like \apiname{sendfile}, the \mynamep{} emulates its functionality to ensure seamless operation while leveraging \mytcp{} offloading.

\section{\mytcp{} Design and Implementation}\label{sec:htcp_overview}

In this section, we provide a detailed explanation of the design of \mytcp{}. As the core component of \myname{}, \mytcp{} is crucial to its functionality and usability. We first outline the design goals of \mytcp{}, followed by an in-depth discussion of its components.

\subsection{\mytcp{} Design Goals}

\mytcp{} is responsible for enabling the execution of the host TCP stack on the DPU, addressing several key challenges. To ensure the usability of \myname{}, \mytcp{} must achieve the following objectives:

\begin{itemize}[leftmargin=*]
\item Efficient execution on SoC platforms, specifically within the ARM subsystem of DPUs, ensuring high scalability through multi-threading to support rapid scaling across the available ARM cores and flexible deployment.

\item  Effective offloading of the host TCP stack while maintaining comparable CPU savings on the host side.

\item  Full compatibility with POSIX socket APIs, ensuring seamless integration as a complete TCP stack that adheres to standard TCP interactions.
\end{itemize}

\subsection{\mytcp{} Overview}

\begin{figure}[ht]
    \centering
    \includegraphics[width=0.45\textwidth]{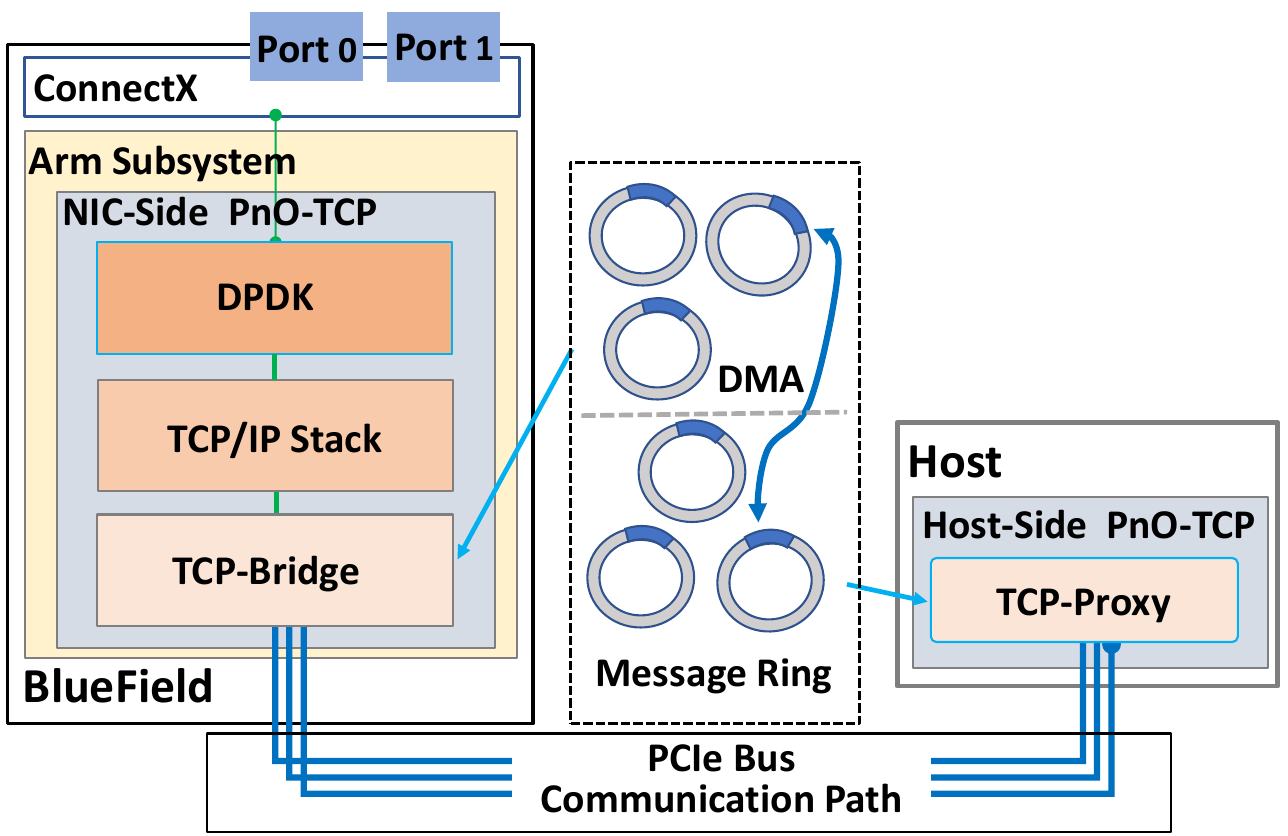}
    \caption{\mytcp{} Architecture: Host-NIC Cooperative TCP Stack}
    \label{fig:htcp_layer}
\end{figure}

Fig.~\ref{fig:htcp_layer} illustrates the \mytcp{} architecture, which is divided into two parts: the host-side \mytcp{} and the NIC-side \mytcp{} on the DPU.

On the host side, \mytcp{} consists of a single component, \mytcph{}, which serves as the API interface layer. It provides POSIX-compliant APIs that can be directly invoked by host network applications. On the DPU, within the ARM subsystem, the NIC-side \mytcp{} comprises three main components: \mytcpn{}, the TCP/IP stack, and DPDK. The \mytcpn{} acts as a bridge between the host-side \mytcph{} and NIC-side TCP/IP stack, forwarding requests and data from \mytcph{} to the TCP/IP stack and returning results back to \mytcph{}. Additionally, the other two components follow a traditional userspace TCP stack structure: the TCP/IP stack implements core protocol logic, while DPDK serves as the network backend for packet transmission.

When a network application initiates a send operation, the process follows these steps:

\begin{enumerate}[leftmargin=*]
    \item The application invokes the API provided by \mytcph{} on the host side.
    \item The operation is transferred to the \mytcpn{} in the ARM subsystem via DMA.
    \item The \mytcpn{} forwards the data to the TCP/IP stack, where it is processed and encapsulated into TCP packets.
    \item The processed data is transmitted through the DPDK layer and sent to the NIC queue.
\end{enumerate}

Throughout this process, all TCP protocol operations including state maintenance, retransmissions, and acknowledgments are executed within the ARM subsystem, eliminating the need for host CPU involvement and achieving significant CPU savings on the host side.

\subsection{\mytcph{} and \mytcpn{}}

The primary function of \mytcph{} and \mytcpn{} is to facilitate communication between the host and the DPU, enabling interaction between applications running on the host and the TCP stack on the DPU. As illustrated in the Fig.~\ref{fig:htcp_layer}, they communicate via Message Rings, which is implemented as a shared memory structure present on both the host and the DPU. Data synchronization is maintained through DMA, ensuring efficient communication across PCIe. 

The key responsibilities of \mytcph{} include: \textcircled{1} Serving as an interface for host applications by providing POSIX-compliant network APIs. \textcircled{2} Managing the host-side \mytcp{} Message Ring by filling requests from host applications into the appropriate Message Ring and delivering data from the Message Ring back to the host applications.

Similarly, \mytcpn{} is responsible for: \textcircled{1} Synchronizing the Message Ring through DMA operations. It performs polling to check whether new data from the host needs to be transferred to the DPU and ensures that updated Message Ring data from the DPU is synchronized back to the host. All DMA operations are initiated by \mytcpn{}. \textcircled{2} Parsing requests received from the host and forwarding them to the TCP/IP stack for further processing.
\newline


\zztitle{Message Ring} To mitigate the DMA latency bottlenecks outlined in Challenges 2 and 3 of Section~\ref{sec:challenges}, we need to minimize both the number of DMA operations and the number of synchronous calls from the host to the DPU (i.e., calls where the host waits for a response from a component on the DPU before returning). Our approach involved two key strategies: first, amortizing DMA overhead by batching multiple requests; second, enabling more calls to return directly to the application on the host side without waiting for a DPU component response (e.g., allowing a write operation to return to the application as soon as the data is in the host-side buffer). Toward these ends, we designed a multi-ring message ring system for efficient data transfer, as illustrated in Figure~\ref{fig:htcp_layer}. We began by classifying socket APIs into two categories: set-type (S-type) and get-type (G-type). S-type APIs encompass operations that send data to the TCP stack or configure TCP-related functions, including \apiname{socket}, \apiname{listen}, \apiname{send}, \apiname{sendto}, \apiname{write}, and \apiname{writev}, among others.  Conversely, G-type APIs retrieve data from the TCP stack, such as \apiname{read}, \apiname{epoll\_wait}, \apiname{recv}, and \apiname{recvfrom}. S-type operations allow batching multiple requests that do require a response from a DPU component. Furthermore, requests such as \apiname{write} and related functions can return immediately without waiting for that response. G-type operations, in contrast, can be completed entirely on the host side, allowing a return to the application without waiting for a DPU component. To optimize data interaction, S-type and G-type APIs utilize distinct message rings: S-type APIs are serviced by an S-type message ring, while G-type APIs leverage a dedicated G-type message ring.

\zztitle{S-type Message Ring} The S-type message ring consists of a single ring shared by all S-type APIs. For each API request, a contiguous block of the required size is sequentially allocated from the ring. Each block is prefixed with an 8-byte header—comprising a 4-byte flag and a 4-byte length field—indicating the block’s attributes and size, respectively, followed by the request-specific data. All blocks are 8-byte aligned and stored contiguously within the ring, enabling a single DMA operation to transfer multiple blocks. This design amortizes DMA latency, reducing the average delay.

\begin{figure}[ht]
    \centering
    \includegraphics[width=0.49\textwidth]{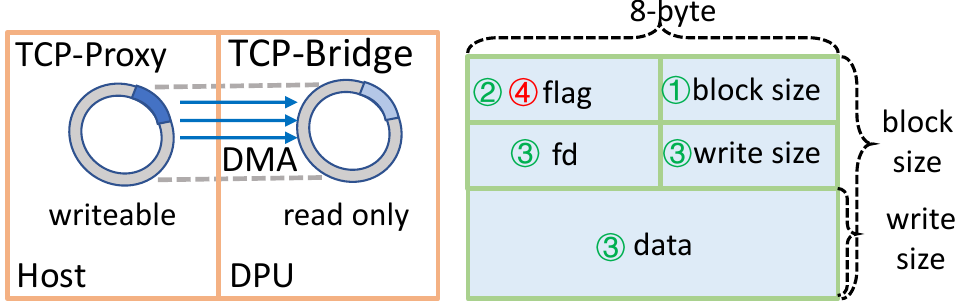}
    \caption{On the left is the S-type Message Ring, and on the right is the write block and the sequence in which the data is updated.}
    \label{fig:s_type_ring}
\end{figure}

To ensure memory consistency between the DPU and the host while allowing multiple threads to operate concurrently, we enforce the following rules: \textcircled{1} Mutual exclusion is required only when allocating blocks. After filling in the block contents, a memory barrier is applied before updating the block flag. When \mytcpn{} detects the updated flag, it ensures that the data has been fully updated by \mytcph{}. \textcircled{2} Only \mytcph{} can request blocks and has write permissions, while \mytcpn{} has read permissions and can modify only flag field.

For \apiname{socket} and other configuration APIs, caller waits synchronously for the return value. It reserves a writable position for \mytcpn{} and busy-waits at the return value's location until it is updated by \mytcpn{}. In contrast, other APIs, such as write operations that submit data into the protocol stack, allow the caller to return immediately after submitting data to the Message Ring. As illustrated in the right of Fig.~\ref{fig:s_type_ring} for a write operation, the process begins with requesting a block of size \apiname{write\_size + 16} by modifying \textcircled{1}\textcircled{2}. The file descriptor (fd), write size, data, and block length are then inserted into the block. To ensure data consistency on the DPU, a memory barrier is enforced before updating the block flag \textcircled{4} to \apiname{W\_WRITE\_FLAG}. In the ARM subsystem, \mytcpn{} continuously polls the Message Ring through DMA synchronization. When \mytcpn{} detects a block flag that is not \apiname{W\_NONE}, it parses the block and performs the corresponding socket operation. 
\newline

\begin{figure}[ht]
    \centering
    \includegraphics[width=0.35\textwidth]{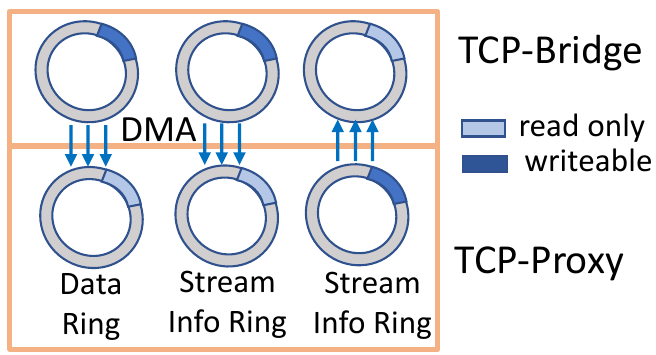}
    \caption{G-type Message Ring.}
    \label{fig:g_type_ring}
\end{figure}

\zztitle{G-type Message Ring} 
The G-type message ring is composed of multiple message rings. As illustrated in Fig.~\ref{fig:g_type_ring}, these include both data rings and two socket stream information rings. The Data Ring is used to transfer data received by the TCP stack that is destined for the \mytcph{}. Specifically, it synchronizes read data and epoll events, with separate data rings dedicated to each type of data. In this architecture, message rings exist in paired configurations, one on the host and one on the DPU—operating under a producer-consumer model. For the Data Ring, the NIC-side \mytcpn{} acts as the producer, while the host-side \mytcph{} is the consumer. Consequently, only the DPU’s \mytcpn{} holds both read and write permissions for the Data Ring, whereas the host-side \mytcph{} has read-only access.

The Stream Info Ring is responsible for conveying socket stream metadata. This metadata includes the location of socket stream data within the Data Ring and the number of epoll events associated with that stream. To prevent overlapping of data within the ring, the Stream Info Ring also maintains pointers to the head and tail positions of its data. Structurally, the Stream Info Ring is comprised of fixed-length blocks, with each block corresponding to a specific socket stream (i.e., a file descriptor). Since both \mytcph{} and \mytcpn{} must update stream information, two separate Stream Info Rings are employed, each component owning its own writable ring to synchronize stream metadata and positional data.


Our measurements indicate that the NIC's DMA engine does not guarantee ordering, meaning that the completion order of DMA operations can deviate significantly from the order in which requests are submitted. To ensure that \mytcph{} maintains consistent memory states despite potential DMA reordering, the Data Rings must be fully synchronized (e.g., ensuring data DMA completion via appropriate barriers) before the corresponding metadata is updated in the Stream Info Rings on the host. This strict ordering is crucial for preventing data inconsistencies during high-throughput operations

For instance, when data for file descriptor 1002 (e.g., stream 2) is received, \mytcpn{} first allocates a block from the Data Ring and copies the incoming data into it. Subsequently, it uses the file descriptor to compute a hash and locate an empty block in the Stream Info Ring, where it populates the corresponding stream metadata. After flushing the Data Ring to the host, \mytcpn{} synchronizes the Stream Info Ring with the host. When \mytcph{} later performs a read operation on fd 1002, it retrieves the associated stream metadata from the Stream Info Ring via hashing, copies the data from the Data Ring into the network application's buffer, and thereby completes the read operation.

Essentially, \mytcph{} serves as a cache for \mytcp{} stream data on the host side, enabling network applications to efficiently execute socket operations. Meanwhile, \mytcpn{} functions as a service layer for \mytcph{}, with data exchanged between these components via ring buffers. To ensure memory consistency, each shared ring queue is configured to be writable only from one side, and synchronization is rigorously maintained through carefully sequenced operations and memory barriers.

\subsection{TCP/IP Stack}

The TCP/IP stack encapsulates the processing logic for the TCP protocol, serving as a foundational component in traditional networking systems. To optimize performance and support offloading of the host TCP stack, we opted against using the Linux kernel TCP stack within the ARM subsystem of the Data Processing Unit (DPU). Instead, we developed a custom user-space TCP protocol stack tailored to our requirements. Several established user-space stacks were evaluated, including F-Stack~\cite{fstack}, Seastar~\cite{seastar}, and mTCP~\cite{jeong2014mtcp}. However, F-Stack employs a multi-process architecture rather than a multi-threaded one, which conflicts with our need for efficient data sharing. Seastar~\cite{seastar}, an asynchronous framework built on C++ libraries, lacks compatibility with the POSIX socket API. While mTCP~\cite{jeong2014mtcp} offers excellent scalability, no community-supported version for ARM was available at the time of development. Consequently, we designed our own user-space TCP stack, drawing on mTCP’s principles and incorporating its core TCP protocol implementations, such as basic TCP processing and timer management.



Given the performance constraints of the ARM core (Challenge 1, Section~\ref{sec:challenges}), optimizing the TCP/IP stack was crucial for achieving high efficiency on the ARM subsystem. Our primary optimization strategy focused on eliminating redundant data copying. Therefore, we implemented a zero-copy design for both data transmission and reception, detailed below.

\begin{figure}[ht]
    \centering
    \includegraphics[width=0.4\textwidth]{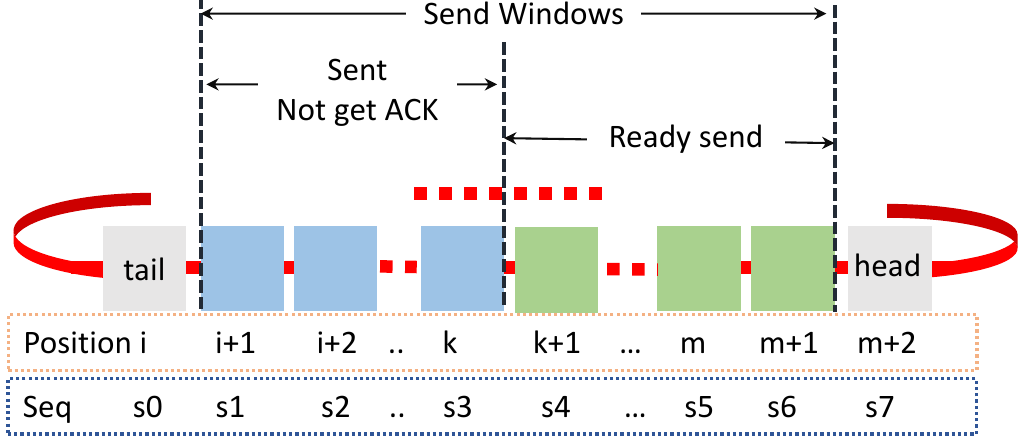}
    \caption{ TCP Send Window Management in \mytcp{}.}
    \label{fig:ring_windows}
\end{figure}

\textbf{Data Transmission.} For outbound data, we pre-allocate a data block slightly larger than the Network Interface Card (NIC) Maximum Transmission Unit (MTU), reserving 128 bytes at the block’s header to accommodate potential protocol stack headers (e.g., TCP, IP, and Ethernet). The payload is then written directly into the data block. As the packet traverses the TCP stack, protocol processing incrementally constructs the headers within the reserved space, assembling the complete packet without additional data copies. The finalized packet is transmitted to the NIC via the Data Plane Development Kit (DPDK). To support TCP retransmission, we simulate the TCP send window using a ring structure, as illustrated in the Fig.~\ref{fig:ring_windows}. The send window dynamically adjusts based on sequence and acknowledgment numbers. For efficient selective retransmission, we maintain a hash table mapping each packet’s sequence number to its position in the ring, enabling rapid lookup and retransmission of specific packets as needed.

\textbf{Data Reception.} On the receiving end, incoming packets are placed directly into a TCP receive pool, managed by a priority queue ordered by sequence numbers. This pool accommodates out-of-order packets, reordering them to reconstruct complete datagrams before transferring them to the reception queue. Packets are retained in memory until \mytcpn{} reads the data from the reception queue, at which point they are released. To optimize resource usage, redundant retransmitted packets detected in the receive pool are automatically discarded. For overlapping packets, we dynamically adjust packet lengths and update TCP headers within the pool to eliminate redundant segments, enhancing both memory efficiency and processing performance.

Throughout the packet handling process, we utilize fixed-size data blocks, slightly larger than the MTU, as the basic unit of data. Each block is identified by its TCP sequence number. By manipulating pointers and modifying headers in place, we achieve protocol packet assembly, retransmission, and reordering without data copying. For additional TCP protocol management features—such as flow control and congestion handling—we adopt established mechanisms from mTCP, leveraging its robust and proven implementation.

\subsection{Thread Management}
On the ARM subsystem, TCP processing involves both DPDK-based polling of the NIC and periodic data exchange with the TCP stack, necessitating efficient coordination among multiple threads. To minimize inter-core contention, our implementation of \mytcp{} adopts mTCP’s lock-free design by localizing resources, such as flow pools and socket buffers, to individual cores. Additionally, we leverage Receive Side Scaling (RSS) to enforce flow-level core affinity, ensuring that packets belonging to the same flow are processed on the same core.

To further optimize cache utilization and mitigate cache migration across physical cores, we refined the threading model originally proposed by mTCP. In mTCP’s default configuration, each CPU core hosts two kernel-level threads: a TCP stack thread, responsible for protocol processing and timer management, and an application thread, enabling independent operation of the TCP logic. However, this approach introduces significant context-switching overhead due to frequent transitions between kernel-level threads. To address this inefficiency, we adapted mTCP by integrating cooperative user-level threading, drawing on the lthread framework~\cite{lthread}. Consequently, \mytcpn{} and the TCP stack now execute as two cooperative user-level threads co-located on the same physical core. This design substantially reduces context-switching costs and minimizes cache pollution, enhancing overall performance on the ARM subsystem.

\subsection{Implementation}

The implementation of \mytcp{} is developed in C. For DMA operations, we utilized the DOCA DMA framework~\cite{doca_dma}. Excluding the code for DPDK and the TCP/IP stack, the total implementation comprises 8,400 lines of code (LOC), with the \mytcph{} component accounting for 2,386 LOC. The DPDK and TCP/IP stack implementations reuse some existing code from mTCP~\cite{jeong2014mtcp} and lthread~\cite{lthread}, and thus their LOC contributions are not separately quantified. The socket API in \mytcp{} prepends the prefix \apiname{ptcp\_} to POSIX socket API names while maintaining full compatibility with POSIX socket definitions.

\section{Evaluation}
To assess the effectiveness of \myname{}, we designed our evaluation around three key objectives. First, we aim to confirm that \mytcp{} correctly executes the TCP stack and seamlessly interacts with standard Linux applications over Linux TCP. Second, we seek to validate that \myname{} can transparently offload the TCP stack of real-world network applications to the DPU without disrupting their functionality. Third, we investigate the performance benefits of \myname{} offloading, particularly its ability to significantly reduce host CPU utilization. To address these objectives, we tested \myname{} with a range of applications: a custom-developed Echo program, Redis (a widely used key-value store), HAProxy (a prominent network traffic load balancer), and Lighttpd (a popular web server).

\subsection{Experiments Setup}

We conducted evaluations on a server equipped with a Mellanox BlueField-3 DPU, featuring 16 ARM A78 cores and an integrated ConnectX-7 NIC. Table~\ref{tab:server_spec} lists the server specifications.

\begin{table}[ht]
    \caption{Specification of the server for evaluation.}
    \label{tab:server_spec}
    \centering
    \begin{tabular}{c|c} \hline
      Host CPU  &  AMD EPYC 7302 \\ \hline
    Host Memory &     128G \\ \hline
    SmartNIC & Mellanox Bluefield-3 DPU\\ \hline
    Host Kernel & Linux 5.15.0-101-generic\\ \hline
    DPU OS & DOCA\_2.2.0\_BSP\_4.2.0\_Ubuntu2204 \\ \hline
    DPU Kernel & Linux 5.15.0-1021-bluefield \\ \hline
    DOCA Version & 2.2.0 \\ \hline
    \end{tabular}

\end{table}

The server is directly connected to a client machine with ConnectX-5 NIC. Both NICs work in default mode. In the NIC-side \mytcp{} backend, we employ DPDK version 21.11. Due to the limited computational resources of the DPU and its additional responsibilities such as traffic monitoring, we adopted a conservative core utilization strategy to demonstrate scalability while ensuring transparency and maintaining a seamless user experience. For each host network thread, we configured a corresponding DPDK thread on the DPU to provide service. Our goal was to reduce host CPU usage while maintaining scalability.

\subsection{Microbenchmark}

\begin{figure}[ht]
    \centering
    \includegraphics[width=0.40\textwidth]{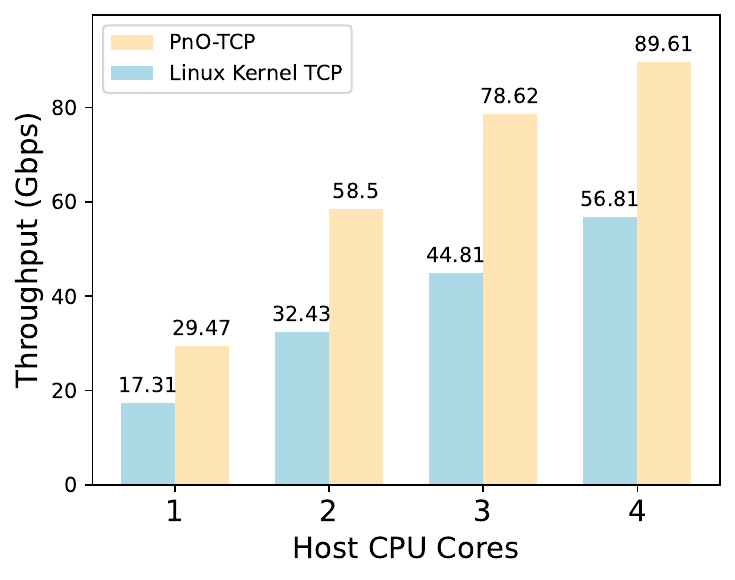}
    \caption{iperf Throughput Performance Comparison of \mytcp{} and Linux Kernel TCP with Varying Number of Host CPU Cores.}
    \label{fig:iperf_tp}
\end{figure}

\zztitle{iperf} To evaluate \mytcp{}'s performance and correctness, we implemented an iperf application using the \mytcp{} API. This application ran on our test host machine, while the client side used normal Linux TCP-based network programs. As shown in the Fig.~\ref{fig:iperf_tp}, in our test environment, iperf with \mytcp{} achieved near line-rate NIC speeds using only 4 host CPU cores. With the same number of cores (\textless 4 cores), the iperf with \mytcp{} performance was approximately 1.7 times faster than the standard implementation. This improvement is primarily attributed to \mytcp{}'s efficient stack protocol based on DPDK. These results demonstrate \mytcp{}'s viability and efficiency, showing it significantly reduces host CPU utilization while improving SmartNIC utilization rates. \newline

\zztitle{Echo} To evaluate the correctness and feasibility of \myname{}, we implemented a Network APP Echo based on the POSIX socket API. The Echo program uses a multiplexed I/O model, where the polling thread uses \apiname{epoll\_wait} to determine if any socket is readable. Then, for each readable socket, it writes any data read from the socket back to the same socket. We ran a client program on the client machine that established 120 TCP connections with the server. Each connection sent a fixed-size packet and waited to receive a packet of the same size before sending the next one.

\begin{figure}[ht]
    \centering
    \includegraphics[width=0.45\textwidth]{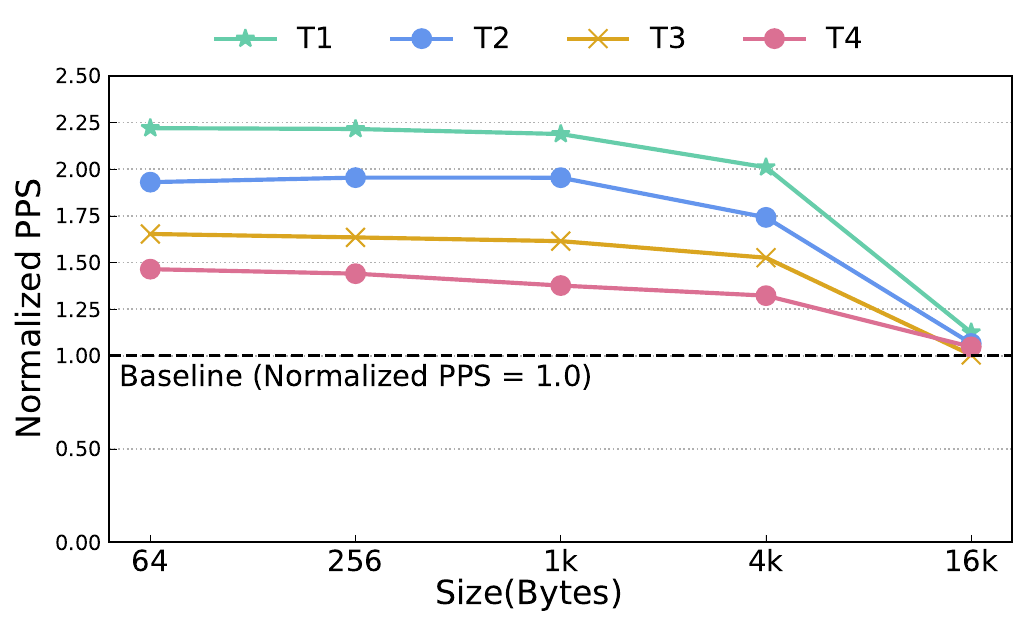}
    \caption{Echo with \myname{} Performance Gain: Echo Normalized PPS Benchmark. T\# means the number of threads.}
    \label{fig:echo}
\end{figure}

As shown in the Fig.~\ref{fig:echo}, we used Linux TCP-based Echo data as our baseline (Normalized PPS = 1) and compared it with the PPS of TCP stack offloaded using \myname{}. The results display the relative PPS sizes under different thread counts. With a single thread, the Echo with \myname{} achieved the maximum performance gain of 2.23 times. However, as the size increased, performance acceleration gradually decreased due to the processing capacity limitations of the \mytcp{} backend. As shown in Table~\ref{tab:cpu_ut}, while performance acceleration decreased, host CPU utilization also declined. Furthermore, as the number of threads increased, the relative performance gain decreased. This occurred because more threads led to increased traffic on the DPU, resulting in greater memory buffer resource usage and DMA resource consumption, which limited thread scalability and prevented the maintenance of consistent performance gains~\cite{Li_Chen_Shen_Wang_Cao_2025}. Although this effect was anticipated, our core objective of conserving host CPU resources was consistently achieved, with the entire TCP execution on DPU saving approximately 56\% of CPU resources. Therefore, to obtain maximum performance benefits, we recommend applying this approach to applications with fewer network threads.

\subsection{Application Benchmark}
We investigated the CPU savings and performance benefits of applying \myname{} to real-world applications. We selected some of the most common network applications: Redis (a key-value store), HAProxy (a widely used load balancer for web traffic), and Lighttpd (a popular HTTP server). \newline

\zztitle{Redis}
We utilized the in-house Redis-benchmark tool as the client to assess the Requests Per Second (RPS) performance of Redis across various request sizes. The benchmark was configured with 200 concurrent connections and 8 threads, adhering to Redis’s default configuration, excluding the specification of I/O thread count. In this default setup, writing responses, identified as a frequent bottleneck that lends itself to parallelization, is handled by the main thread, which partitions responses into batches and distributes them among worker threads, including itself. However, in the default configuration, read operations are not delegated to I/O threads and are instead processed directly by the main thread. We use Redis in 6.2.6 version with all default configurations in our experiments. 


\begin{figure*}[ht]
    \centering
    \subfloat[\apiname{Redis GET}]{
        \includegraphics[width=0.32\textwidth]{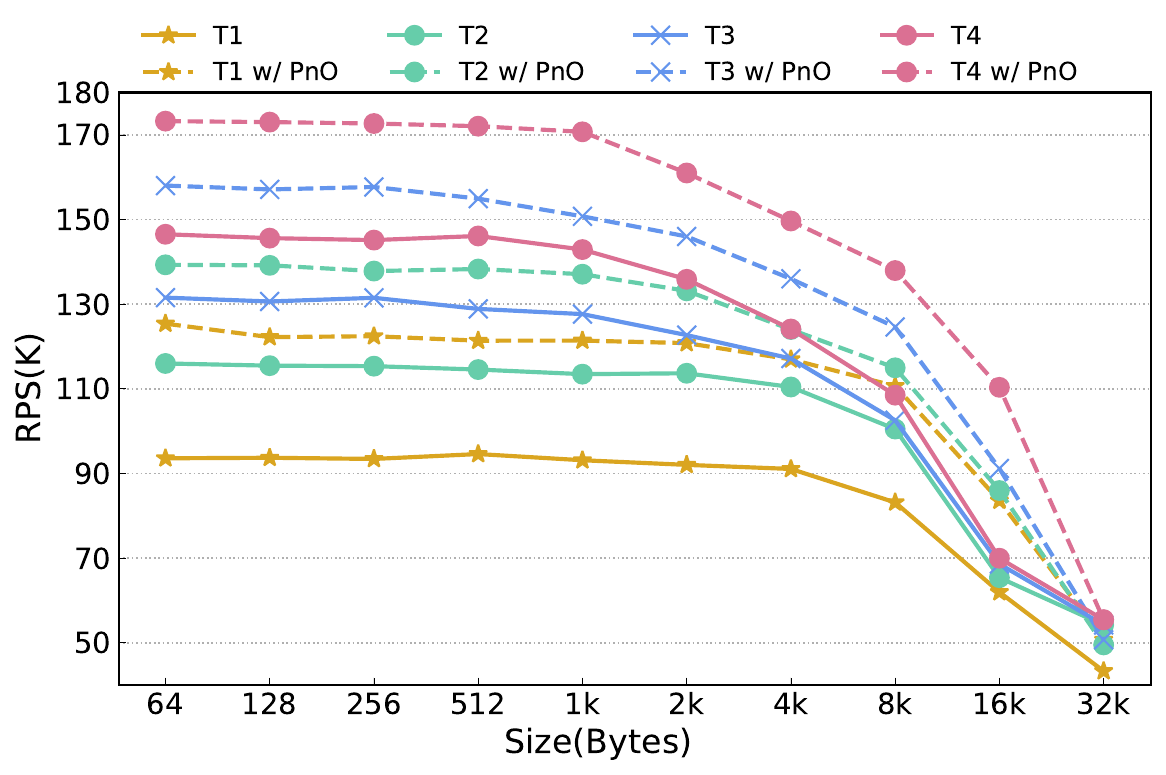}
        \label{fig:redis_get}
    }
    \subfloat[\apiname{Redis SET}]{
        \includegraphics[width=0.34\textwidth]{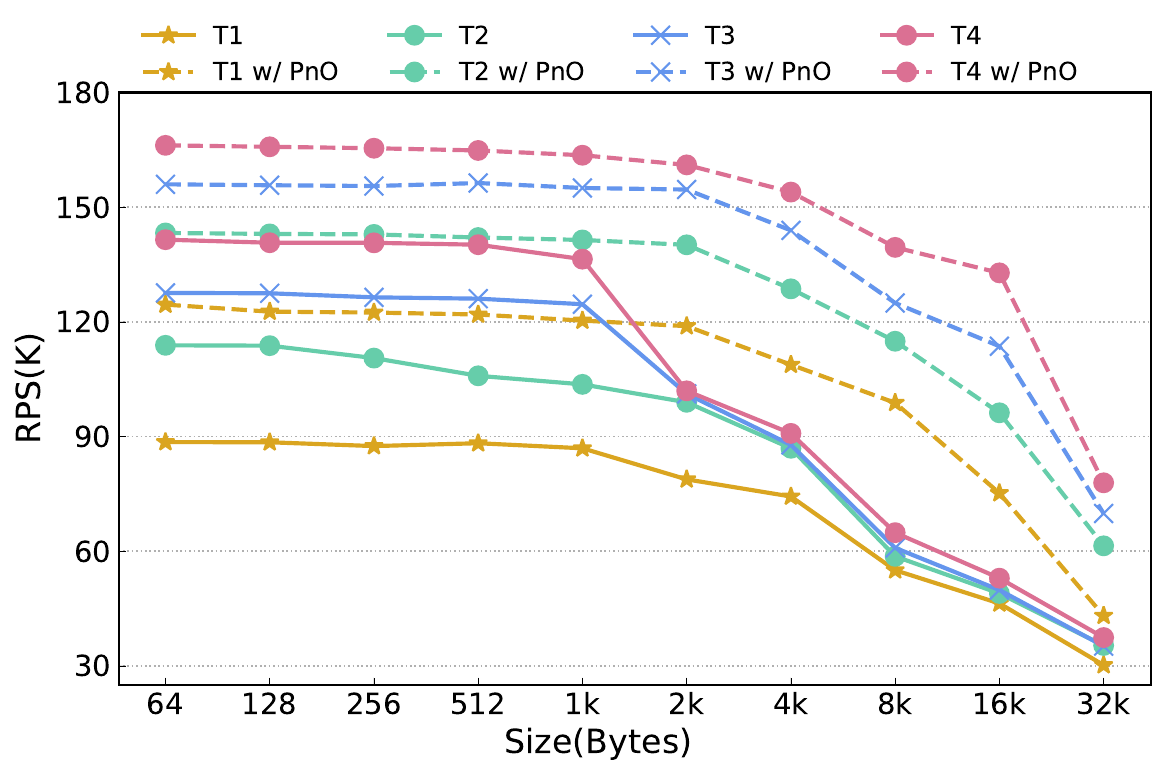}
        \label{fig:redis_set}
    }
    \subfloat[\apiname{Lighttpd}]{
        \includegraphics[width=0.34\textwidth]{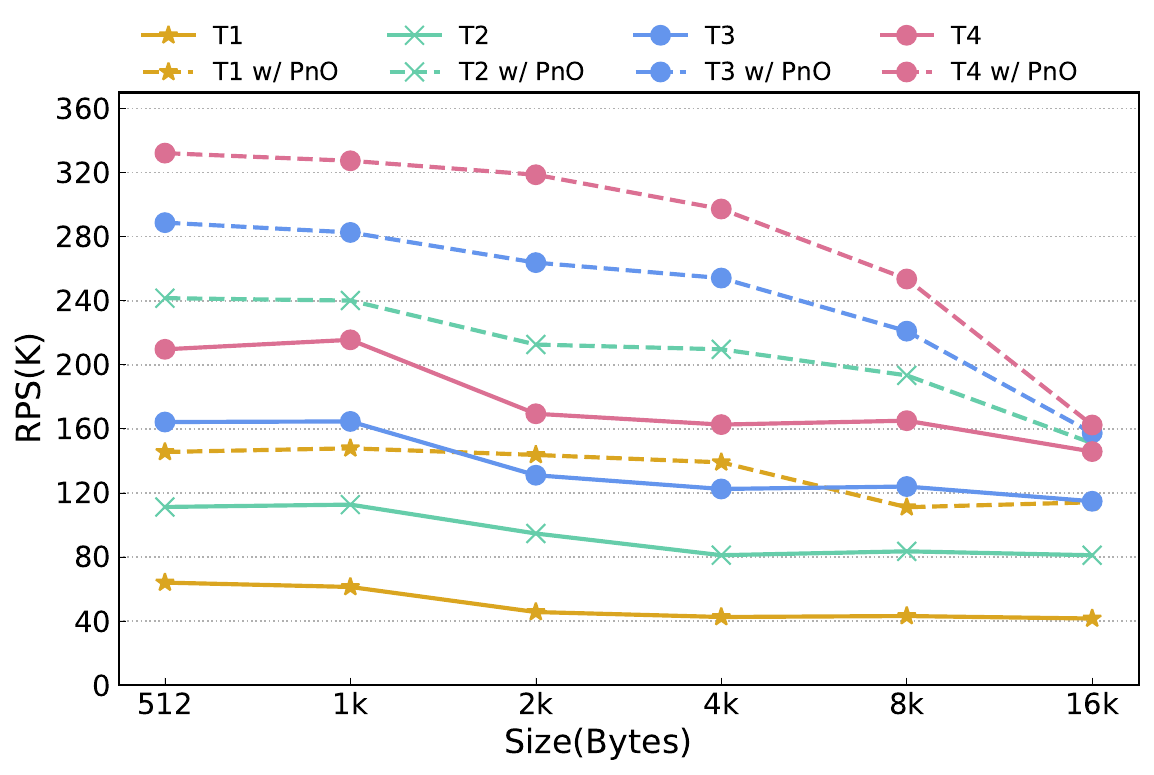}
        \label{fig:lighttpd}
    }

    \caption{ (a). Performance of \apiname{Redis GET}.~~~~(b). Performance of \apiname{Redis SET}.~~~~(c). Performance of \apiname{Lighttpd} against different number of threads. (T\# means the number of threads.)}
    \label{fig:second}
\end{figure*}

\begin{figure*}[ht]
    \centering
    \subfloat[\apiname{Lighttpd}]{
        \includegraphics[width=0.31\textwidth]{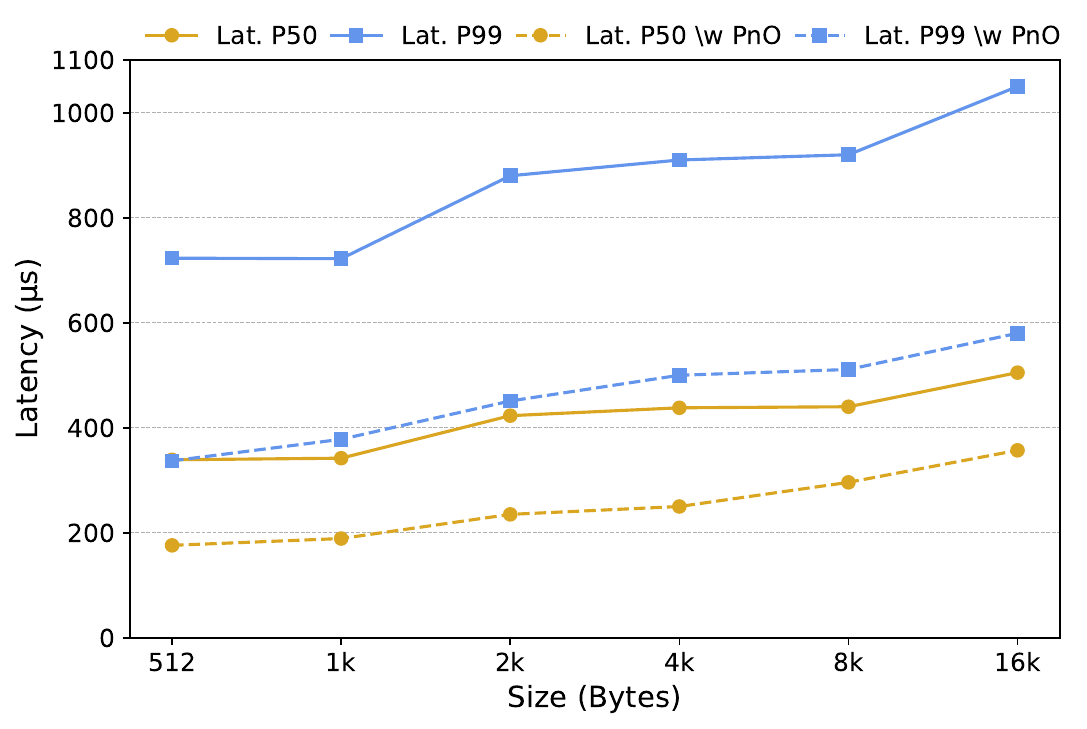}
        \label{fig:p99_lhttpd}
    }
    \subfloat[\apiname{Lighttpd}]{
        \includegraphics[width=0.32\textwidth]{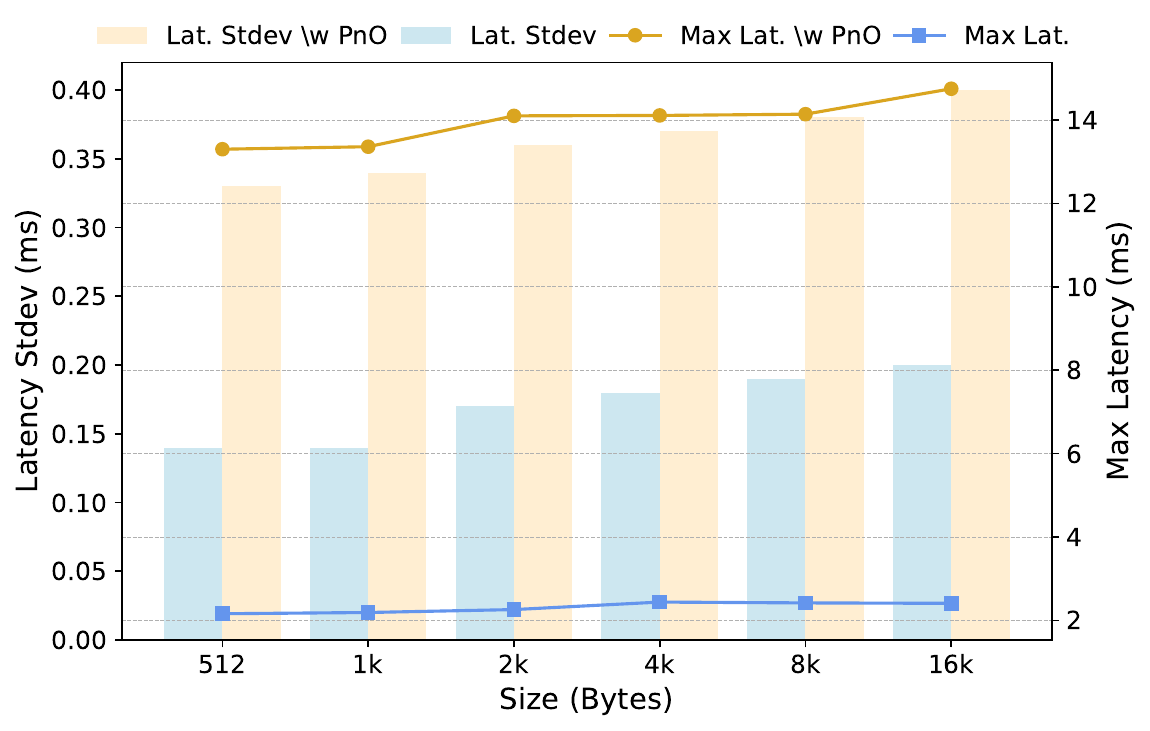}
        \label{fig:lat_lhttpd}
    }
    \subfloat[\apiname{HAProxy}]{
        \includegraphics[width=0.32\textwidth]{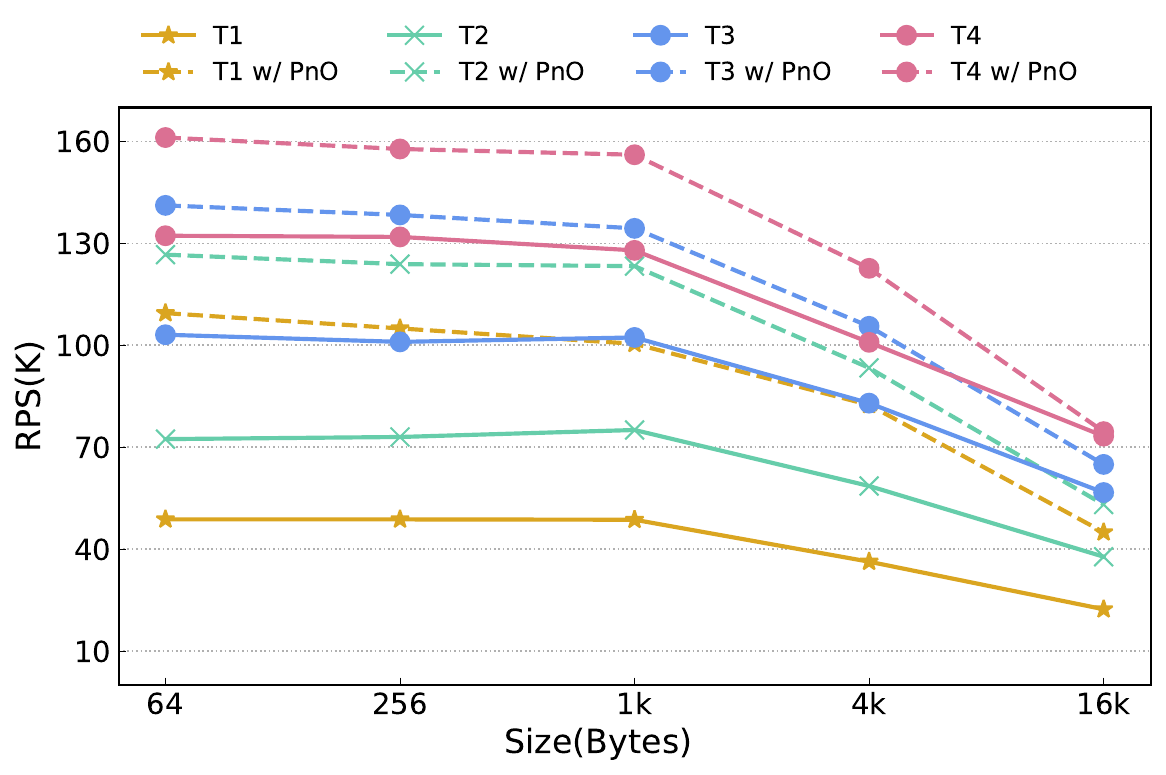}
        \label{fig:haproxy_off_stack}
    }
    \caption{ (a). p50 and p99 Latency  of \apiname{Lighttpd} against the different sizes.~~~~(b). Latency standard deviation and Max Latency of \apiname{Lighttpd} against the different sizes.~~~~(c). Performance of \apiname{HAProxy} against thread count. T\# means the number of threads.}
    \label{fig:second}
\end{figure*}

As shown in the Fig.~\ref{fig:redis_get}, we measured the performance of Redis GET operations across varying numbers of network threads and different value sizes, expressed in RPS. In the figure, T1 denotes execution with only the main thread, without additional I/O threads, whereas T2 represents one additional I/O thread, and so on. The solid lines indicate the Linux TCP baseline performance. We observed that as the number of threads increased, the baseline RPS did not scale proportionally due to the constraint of a single main thread. The dashed lines depict the RPS results under \myname{}, demonstrating performance improvements for value sizes smaller than 32 KB. Similar to our previous Echo experiment, when the value size exceeded 32 KB, no noticeable RPS gains were observed. Since I/O threads in Redis under \myname{} are limited to handling write operations, CPU utilization on the host remained unchanged, as Redis did not exceed a single CPU core.

To validate this observation, we compared two scenarios: one where only one thread executes on the host (T1 in the Fig.~\ref{fig:redis_get} and Fig.~\ref{fig:redis_set}) and another with \myname{} (T1 w/ \myname{}). For small data sizes (64 bytes), the RPS increased from 93K to 125K, indicating a 34\% performance improvement by leveraging NIC offloading. However, as the data size increased to 2 KB, the performance gain diminished. Our analysis attributes this behavior to the NIC Maximum Transmission Unit (MTU). When the request size is small, it fits within a single packet, leading to efficient processing. Conversely, when the request size exceeds 1500 bytes, multiple packets are required, increasing processing overhead and reducing the overall RPS improvement.


As depicted in the Fig.~\ref{fig:redis_set}, SET operations exhibit similar performance trends as GET operations. Redis SET with \myname{} achieves performance comparable to Redis GET with \myname{}. However, the Redis SET baseline more clearly highlights the bottleneck caused by the main thread, as all read operations are managed by the main thread. Consequently, adding I/O threads yields only limited improvements in SET RPS.

Finally, in Redis, \apiname{epoll\_wait} monitors not only network events but also snapshot-related pipeline events, demonstrating the robustness of \mytcp{} across diverse application scenarios. \newline

\zztitle{Lighttpd}
We evaluate the multi-threaded Lighttpd used by mTCP~\cite{jeong2014mtcp}. Similar to HAProxy, Lighttpd also follows epoll-read-write thread model. We run wrk~\cite{wrk} as the client with 8 threads and  100 concurrent connections.



As the Fig.~\ref{fig:lighttpd} shows, the RPS gains obtained were similar to our previous experiments, successfully achieving our offloading objectives. Specifically, in the single-thread, 512-byte scenario, we observed a performance improvement of 127\%, with RPS increasing from 63.9K to 145.5K. Still, offloading yields good acceleration for all cases.



To assess \myname{}'s impact on traffic quality, we examined the latency characteristics in Lighttpd under a single-threaded configuration. As shown in the Fig.~\ref{fig:p99_lhttpd}, both p50 and p99 latencies are lower with \myname{} compared to the baseline. This reduction is primarily attributed to \myname{}'s direct packet processing approach, which bypasses operating system involvement and eliminates the overhead of system calls present in a host-based implementation. Furthermore, we evaluated the standard deviation of request latency and the maximum latency in Lighttpd, comparing \myname{}'s implementation against the baseline (Fig.~\ref{fig:lat_lhttpd}). While \myname{} enhances throughput, it increases latency variation, with the standard deviation rising from 0.17 ms to 0.35 ms and the maximum latency increasing from 2 ms to 14 ms (Fig.~\ref{fig:lat_lhttpd}). This aligns with our expectations, as batch processing, used to amortize DMA latency, introduces variability by grouping requests of slightly different arrival times.



\zztitle{HAProxy}
HAProxy is a widely used load balancer for web traffic. It follows a basic epoll-read-write thread model and does not use a polling thread.  We use the in-house benchmark tool \apiname{dpbench}~\cite{dpbench} as the client with 80 HTTP connections and 8 threads.


For this experiment, we configured a Layer-7 HTTP load balancer, with clients using \apiname{h1load} from \apiname{dpbench}. As shown in the Fig.~\ref{fig:haproxy_off_stack}, HAProxy with \myname{} demonstrates performance benefits consistent with our previous results. As traffic volume increases, the limitations of \myname{} become apparent and the performance gains decrease. However, during this process, CPU utilization is significantly reduced. \newline

\subsection{CPU Utilization}
In this section, we analyze the CPU cycle saving for all the apps. Table~\ref{tab:cpu_ut} shows the CPU utilization for all the experiments. 

\begin{table}[ht]
    \centering
        \caption{CPU Utilization. T\# means the number of threads.}
    \label{tab:cpu_ut}
    \resizebox{0.48\textwidth}{!}{
    \begin{tabular}{c|cccc}
        \hline
       App & T1 & T2 & T3 & T4 \\
        \hline \hline
         Echo & 85\%--102\% & 198\%--202\% & 298\%--303\% & 397\%--405\% \\
        Echo w/ \myname{} & 95\%--100\% & 105\%--125\% &140\%--160\%  &155\%--195\%  \\
        \hline \hline
        Redis & 99\%--101\% & 190\%--202\% & 290\%--302\% & 385\%--402\% \\
        Redis w/ \myname{}  & 72\%--100\% & 72\%--100\% & 72\%--100\% & 72\%--100\% \\
        \hline \hline
        HAProxy & 99\%--100\% & 198\%--202\% & 298\%--300\% & 397\%--399\% \\
        HAProxy w/ \myname{} & 95\%-100\% &95\%--120\%  & 95\%--140\%& 96\%--165\%\\
        \hline \hline
        Lighttpd & 97\%--100\% & 196\%--200\% & 295\%--301\% & 390\%--400\% \\
        Lighttpd w/ \myname{} & 69\%--84\% & 138\%--147\% & 160\%--208\% & 204\%--297\% \\
        \hline \hline
    \end{tabular}
    }   
\end{table}


\zztitle{Redis}
In a default Redis configuration, both the main thread and I/O threads each consume nearly a full x86 CPU core. However, Redis's default behavior utilizes I/O threads only for \apiname{write} operations, \apiname{read} operations are handled by the main thread. With \myname{}, the work of these write-focused I/O threads is offloaded to the DPU. Consequently, under \myname{}, host CPU utilization is significantly reduced, as essentially only the main thread remains active on the host, consuming approximately one CPU core.

\zztitle{Lighttpd}
Unlike the other applications tested, Lighttpd has a relatively high proportion of CPU time dedicated to application logic. Consequently, while \myname{} still reduces host CPU utilization by offloading the TCP stack, the overall CPU savings are less dramatic compared to applications with a higher network processing overhead. Despite this, \myname{} still delivers a noticeable improvement in RPS.

\zztitle{HAProxy}
In a traditional setup, each \apiname{HAProxy} thread consumes nearly one CPU core. In contrast, across multi-threaded scenarios, we observed an average host CPU utilization savings of about 50\% when using \myname{} to offload the TCP stack.

\section{Discussion and Future Work}

Our study introduces \myname{}, a novel framework demonstrating the feasibility of transparently offloading the complete TCP stack to DPUs, significantly reducing host CPU utilization.  The core of \myname{} is \mytcp{}, a rigorously engineered, lightweight, userspace TCP stack built upon DPDK.  This design allows \mytcp{} to span both host and DPU, seamlessly integrating communication via optimized message rings. While \myname{} achieves its primary goal of transparent offloading and demonstrates performance improvements, particularly for small packet scenarios, several limitations and opportunities for future research have been identified.

\subsection{DPU Resource Considerations}

While \myname{} significantly reduces host CPU utilization, it's important to acknowledge the resource constraints on the DPU. The DPU's general-purpose CPU cores are not solely dedicated to TCP processing, they also handle other tasks, such as network monitoring and control plane operations. Furthermore, dedicating all DPU cores to accelerating a single host thread would be inefficient and unscalable.

To address this, our current implementation of \myname{} employs a conservative core allocation strategy on the DPU. We use a number of DPU threads equal to the number of network threads on the host, plus one additional thread dedicated to DMA polling. This ensures fairness and prevents the DPU from becoming a bottleneck. Despite this conservative approach, our experimental results demonstrate that the efficiency of \mytcp{} allows this configuration to adequately handle the network load of typical host applications while significantly reducing host CPU usage.

\subsection{Network Latency Jitter}

While \mytcp{} demonstrates throughput gains, experiments with Lighttpd highlight a trade-off: increased network latency jitter, evidenced by higher standard deviation and worst-case latency figures compared to the baseline. A key contributor is the batch processing strategy implemented to optimize host-DPU communication over PCIe. This approach inherently mixes requests from slightly different arrival times within a single batch, disrupting strict FIFO handling. Latency variability is further compounded by the unpredictable nature of DMA completion times and the alternating execution model of our single-threaded, coroutine-based design for DPDK and TCP processing. Crucially, this increased jitter primarily affects latency outliers. Our results show that p99 latency is significantly improved with \mytcp{} compared to the baseline. Moreover, the overall standard deviation increase (around 0.35 ms) represents a noticeable increase in variability relative to low-latency environments where baseline RTTs can be in the millisecond range. However, this increase must be weighed against the achieved CPU savings. In scenarios where host efficiency is paramount, this jitter might be an acceptable trade-off.


\subsection{DPU Memory Bandwidth}
Unlike traditional host environments, DPUs lack Data Direct I/O (DDIO) and are equipped with smaller Last-Level Caches (LLCs). As a result, while CPU resource savings are achieved, performance improvements diminish when handling larger volumes of data and higher memory demands. However, we anticipate that these hardware limitations will gradually be mitigated as DPU technology evolves, with forthcoming generations expected to offer increased memory bandwidth, a greater number of CPU cores, and expanded LLC capacities. Particularly, the adoption of technologies like ARM's Cache Coherent Interconnect (CCI)~\cite{arm_cci400} within DPU SoCs could significantly alleviate memory bandwidth constraints for packet processing, potentially removing it as a major bottleneck in future DPU generations.

\subsection{Future Work}

Based on the successful demonstration of \myname{} for transparent TCP offloading, our future research roadmap includes several key initiatives:

First, we will focus on optimizing the core offloading mechanism. This involves tackling the host-DPU communication bottleneck by exploring next-generation interconnects like CXL to reduce latency and increase bandwidth, potentially mitigating the PCIe hop overhead. We will also refine DMA transfer strategies, investigating adaptive batching and finer-grained controls for improved efficiency.

Second, building on the optimized foundation, we aim to extend the scope of what is offloaded. Beyond the network stack, we plan to investigate offloading application-level logic, such as entire I/O processing threads in applications like Redis, directly onto the DPU. This promises further reductions in host resource consumption and potential performance gains.

Third, we seek to broaden the applicability of \myname{}. This includes adapting the framework for diverse SmartNIC architectures, notably FPGAs, which requires addressing challenges in instruction set compatibility and programming models. Furthermore, we will work towards simplifying deployment and integration by exploring options like kernel module development or API extensions within standard operating systems and networking frameworks.

\section{Conclusion}
\label{sec:conclusion}
In this paper, we present \myname{}, a transparent offloading framework that enables full TCP stack execution on off-path SmartNICs without modifying existing applications, by leveraging \mytcp{}, a high-performance user-space TCP stack optimized for both host and off-path SmartNIC environments. Our approach, which utilizes dynamic API redirection and an efficient message ring design, significantly reduces host CPU utilization and enhances performance, as demonstrated through evaluations with real-world applications such as Redis, HAProxy, and Lighttpd. Despite challenges like PCIe DMA latency and the limited computational power of SmartNIC cores, our results demonstrate the feasibility of transparent TCP offloading, while future work will focus on further minimizing host-NIC interaction overhead and exploring hardware-software co-designs for broader applications.

\bibliographystyle{IEEEtran}
\bibliography{IEEEabrv,main_bib}

\begin{thebibliography}{10}
\providecommand{\url}[1]{#1}
\csname url@samestyle\endcsname
\providecommand{\newblock}{\relax}
\providecommand{\bibinfo}[2]{#2}
\providecommand{\BIBentrySTDinterwordspacing}{\spaceskip=0pt\relax}
\providecommand{\BIBentryALTinterwordstretchfactor}{4}
\providecommand{\BIBentryALTinterwordspacing}{\spaceskip=\fontdimen2\font plus
\BIBentryALTinterwordstretchfactor\fontdimen3\font minus
  \fontdimen4\font\relax}
\providecommand{\BIBforeignlanguage}[2]{{%
\expandafter\ifx\csname l@#1\endcsname\relax
\typeout{** WARNING: IEEEtran.bst: No hyphenation pattern has been}%
\typeout{** loaded for the language `#1'. Using the pattern for}%
\typeout{** the default language instead.}%
\else
\language=\csname l@#1\endcsname
\fi
#2}}
\providecommand{\BIBdecl}{\relax}
\BIBdecl

\bibitem{firestone2018azure}
D.~Firestone, A.~Putnam, S.~Mundkur, D.~Chiou, A.~Dabagh, M.~Andrewartha,
  H.~Angepat, V.~Bhanu, A.~Caulfield, E.~Chung \emph{et~al.}, ``Azure
  accelerated networking:$\{$SmartNICs$\}$ in the public cloud,'' in \emph{15th
  USENIX Symposium on Networked Systems Design and Implementation (NSDI 18)},
  2018, pp. 51--66.

\bibitem{agilio_smartnic}
``agilio-smartnic,'' \url{https://netronome.com/agilio-smartnics/}.

\bibitem{NVIDIA_BlueField2_Datasheet}
{NVIDIA Corporation}, ``Nvidia bluefield-2 dpu datasheet,''
  \url{https://www.nvidia.com/content/dam/en-zz/Solutions/Data-Center/documents/datasheet-nvidia-bluefield-2-dpu.pdf},
  Tech. Rep.

\bibitem{what-is-a-smartnic}
``what-is-a-smartnic,'' \url{https://www.xelera.io/post/what-is-a-smartnic}.

\bibitem{NVIDIA_BlueField3_Datasheet}
{NVIDIA}, ``Nvidia bluefield-3 dpu datasheet,''
  \url{https://www.nvidia.com/content/dam/en-zz/Solutions/Data-Center/documents/datasheet-nvidia-bluefield-3-dpu.pdf},
  Tech. Rep.

\bibitem{Broadcom_P225P}
{Broadcom Inc.}, ``P225p - dual-port 25gbe pcie nic with ocp 3.0,''
  \url{https://www.broadcom.com/products/ethernet-connectivity/network-adapters/p225p}.

\bibitem{lin2020panic}
J.~Lin, K.~Patel, B.~E. Stephens, A.~Sivaraman, and A.~Akella, ``$\{$PANIC$\}$:
  A $\{$High-Performance$\}$ programmable $\{$NIC$\}$ for multi-tenant
  networks,'' in \emph{14th USENIX Symposium on Operating Systems Design and
  Implementation (OSDI 20)}, 2020, pp. 243--259.

\bibitem{li2017kv}
B.~Li, Z.~Ruan, W.~Xiao, Y.~Lu, Y.~Xiong, A.~Putnam, E.~Chen, and L.~Zhang,
  ``Kv-direct: High-performance in-memory key-value store with programmable
  nic,'' in \emph{Proceedings of the 26th Symposium on Operating Systems
  Principles}, 2017, pp. 137--152.

\bibitem{kaufmann2016high}
A.~Kaufmann, S.~Peter, N.~K. Sharma, T.~Anderson, and A.~Krishnamurthy, ``High
  performance packet processing with flexnic,'' in \emph{Proceedings of the
  Twenty-First International Conference on Architectural Support for
  Programming Languages and Operating Systems}, 2016, pp. 67--81.

\bibitem{moon2020acceltcp}
Y.~Moon, S.~Lee, M.~A. Jamshed, and K.~Park, ``$\{$AccelTCP$\}$: Accelerating
  network applications with stateful $\{$TCP$\}$ offloading,'' in \emph{17th
  USENIX Symposium on Networked Systems Design and Implementation (NSDI 20)},
  2020, pp. 77--92.

\bibitem{shinde2013we}
P.~Shinde, A.~Kaufmann, T.~Roscoe, and S.~Kaestle, ``We need to talk about
  $\{$NICs$\}$,'' in \emph{14th Workshop on Hot Topics in Operating Systems
  (HotOS XIV)}, 2013.

\bibitem{chiang2010full}
H.-C. Chiang, Y.-P. Dai, and C.-Y. Wang, ``Full hardware based tcp/ip traffic
  offload engine (toe) device and the method thereof,'' Jan.~12 2010, uS Patent
  7,647,416.

\bibitem{wu2006design}
Z.-Z. Wu and H.-C. Chen, ``Design and implementation of tcp/ip offload engine
  system over gigabit ethernet,'' in \emph{Proceedings of 15th International
  Conference on Computer Communications and Networks}.\hskip 1em plus 0.5em
  minus 0.4em\relax IEEE, 2006, pp. 245--250.

\bibitem{linuxtoe}
T.~L. Foundation., ``Toe,''
  \url{https://wiki.linuxfoundation.org/networking/toe}.

\bibitem{Danielnsdi18}
\BIBentryALTinterwordspacing
D.~Firestone, A.~Putnam, S.~Mundkur, D.~Chiou, A.~Dabagh, M.~Andrewartha,
  H.~Angepat, V.~Bhanu, A.~Caulfield, E.~Chung, H.~K. Chandrappa,
  S.~Chaturmohta, M.~Humphrey, J.~Lavier, N.~Lam, F.~Liu, K.~Ovtcharov,
  J.~Padhye, G.~Popuri, S.~Raindel, T.~Sapre, M.~Shaw, G.~Silva, M.~Sivakumar,
  N.~Srivastava, A.~Verma, Q.~Zuhair, D.~Bansal, D.~Burger, K.~Vaid, D.~A.
  Maltz, and A.~Greenberg, ``Azure accelerated networking: {SmartNICs} in the
  public cloud,'' in \emph{15th USENIX Symposium on Networked Systems Design
  and Implementation (NSDI 18)}.\hskip 1em plus 0.5em minus 0.4em\relax Renton,
  WA: USENIX Association, Apr. 2018, pp. 51--66. [Online]. Available:
  \url{https://www.usenix.org/conference/nsdi18/presentation/firestone}
\BIBentrySTDinterwordspacing

\bibitem{Martysosp19snap}
\BIBentryALTinterwordspacing
M.~Marty, M.~de~Kruijf, J.~Adriaens, C.~Alfeld, S.~Bauer, C.~Contavalli,
  M.~Dalton, N.~Dukkipati, W.~C. Evans, S.~Gribble, N.~Kidd, R.~Kononov,
  G.~Kumar, C.~Mauer, E.~Musick, L.~Olson, E.~Rubow, M.~Ryan, K.~Springborn,
  P.~Turner, V.~Valancius, X.~Wang, and A.~Vahdat, ``Snap: a microkernel
  approach to host networking,'' in \emph{Proceedings of the 27th ACM Symposium
  on Operating Systems Principles}, ser. SOSP '19.\hskip 1em plus 0.5em minus
  0.4em\relax New York, NY, USA: Association for Computing Machinery, 2019, p.
  399–413. [Online]. Available: \url{https://doi.org/10.1145/3341301.3359657}
\BIBentrySTDinterwordspacing

\bibitem{mogul2003tcp}
J.~C. Mogul, ``$\{$TCP$\}$ offload is a dumb idea whose time has come,'' in
  \emph{9th Workshop on Hot Topics in Operating Systems (HotOS IX)}, 2003.

\bibitem{arashloo2020enabling}
M.~T. Arashloo, A.~Lavrov, M.~Ghobadi, J.~Rexford, D.~Walker, and D.~Wentzlaff,
  ``Enabling programmable transport protocols in
  $\{$High-Speed$\}$$\{$NICs$\}$,'' in \emph{17th USENIX Symposium on Networked
  Systems Design and Implementation (NSDI 20)}, 2020, pp. 93--109.

\bibitem{Rajath2019FlexTOE}
\BIBentryALTinterwordspacing
R.~Shashidhara, T.~Stamler, A.~Kaufmann, and S.~Peter, ``{FlexTOE}: Flexible
  {TCP} offload with {Fine-Grained} parallelism,'' in \emph{19th USENIX
  Symposium on Networked Systems Design and Implementation (NSDI 22)}.\hskip
  1em plus 0.5em minus 0.4em\relax Renton, WA: USENIX Association, Apr. 2022,
  pp. 87--102. [Online]. Available:
  \url{https://www.usenix.org/conference/nsdi22/presentation/shashidhara}
\BIBentrySTDinterwordspacing

\bibitem{rfc793}
``Rfc793,'' \url{https://datatracker.ietf.org/doc/html/rfc793}.

\bibitem{dpu_smartnic_market}
\BIBentryALTinterwordspacing
DataHorizzon, ``Dpu smartnlc market by type, by application, and by
  distribution channel, global market size, share, growth, trends, statistics
  analysis report, by region, and segment forecasts 2025-2033.'' [Online].
  Available: \url{https://datahorizzonresearch.com/dpu-smartnic-market-18993}
\BIBentrySTDinterwordspacing

\bibitem{10.1145/3282307}
\BIBentryALTinterwordspacing
J.~L. Hennessy and D.~A. Patterson, ``A new golden age for computer
  architecture,'' \emph{Commun. ACM}, vol.~62, no.~2, p. 48–60, Jan. 2019.
  [Online]. Available: \url{https://doi.org/10.1145/3282307}
\BIBentrySTDinterwordspacing

\bibitem{Pismenny_Eran_Yehezkel_Liss_Morrison_Tsafrir_2021}
\BIBentryALTinterwordspacing
B.~Pismenny, H.~Eran, A.~Yehezkel, L.~Liss, A.~Morrison, and D.~Tsafrir,
  ``Autonomous nic offloads,'' in \emph{Proceedings of the 26th ACM
  International Conference on Architectural Support for Programming Languages
  and Operating Systems}, ser. ASPLOS ’21.\hskip 1em plus 0.5em minus
  0.4em\relax New York, NY, USA: Association for Computing Machinery, Apr.
  2021, p. 18–35. [Online]. Available:
  \url{https://dl.acm.org/doi/10.1145/3445814.3446732}
\BIBentrySTDinterwordspacing

\bibitem{Li_Kashyap_Guo_Lu_2024}
\BIBentryALTinterwordspacing
Y.~Li, A.~Kashyap, Y.~Guo, and X.~Lu, ``Compression analysis for bluefield-2/-3
  data processing units: Lossy and lossless perspectives,'' \emph{IEEE Micro},
  vol.~44, no.~2, p. 8–19, Mar. 2024. [Online]. Available:
  \url{https://ieeexplore.ieee.org/abstract/document/10364358}
\BIBentrySTDinterwordspacing

\bibitem{Marvell_OCTEON_10_Datasheet}
{Marvell}, ``{Marvell OCTEON 10 DPU Platform Product Brief},''
  \url{https://www.marvell.com/content/dam/marvell/en/public-collateral/embedded-processors/marvell-octeon-10-dpu-platform-product-brief.pdf},
  Tech. Rep.

\bibitem{Suresh_Michalowicz_Ramesh_Contini_Yao_Xu_Shafi_Subramoni_Panda_2023}
\BIBentryALTinterwordspacing
K.~K. Suresh, B.~Michalowicz, B.~Ramesh, N.~Contini, J.~Yao, S.~Xu, A.~Shafi,
  H.~Subramoni, and D.~Panda, ``A novel framework for efficient offloading of
  communication operations to bluefield smartnics,'' in \emph{2023 IEEE
  International Parallel and Distributed Processing Symposium (IPDPS)}, May
  2023, p. 123–133. [Online]. Available:
  \url{https://ieeexplore.ieee.org/abstract/document/10177448}
\BIBentrySTDinterwordspacing

\bibitem{Kim_Ng_Gong_Kwon_Yu_Park_2023}
\BIBentryALTinterwordspacing
T.~Kim, D.~M. Ng, J.~Gong, Y.~Kwon, M.~Yu, and K.~Park,
  ``\BIBforeignlanguage{en}{Rearchitecting the {TCP} stack for {I/O-Offloaded}
  content delivery},'' 2023, p. 275–292. [Online]. Available:
  \url{https://www.usenix.org/conference/nsdi23/presentation/kim-taehyun}
\BIBentrySTDinterwordspacing

\bibitem{kim2006connection}
H.-y. Kim and S.~Rixner, ``Connection handoff policies for tcp offload network
  interfaces,'' in \emph{Proceedings of the 7th symposium on Operating systems
  design and implementation}, 2006, pp. 293--306.

\bibitem{pratt2001arsenic}
I.~Pratt and K.~Fraser, ``Arsenic: A user-accessible gigabit ethernet
  interface,'' in \emph{Proceedings IEEE INFOCOM 2001. Conference on Computer
  Communications. Twentieth Annual Joint Conference of the IEEE Computer and
  Communications Society (Cat. No. 01CH37213)}, vol.~1.\hskip 1em plus 0.5em
  minus 0.4em\relax IEEE, 2001, pp. 67--76.

\bibitem{putnam2014reconfigurable}
A.~Putnam, A.~M. Caulfield, E.~S. Chung, D.~Chiou, K.~Constantinides, J.~Demme,
  H.~Esmaeilzadeh, J.~Fowers, G.~P. Gopal, J.~Gray \emph{et~al.}, ``A
  reconfigurable fabric for accelerating large-scale datacenter services,''
  \emph{ACM SIGARCH Computer Architecture News}, vol.~42, no.~3, pp. 13--24,
  2014.

\bibitem{LWN_Corbet_Kernel_Module_Loading}
``Linux and tcp offload engines,'' \url{https://lwn.net/Articles/148697/}.

\bibitem{pismenny2016tls}
B.~Pismenny, I.~Lesokhin, L.~Liss, and H.~Eran, ``Tls offload to network
  devices,'' in \emph{The Technical Conference on Linux Networking (Netdev)},
  2016.

\bibitem{kim2020case}
D.~Kim, S.~Lee, and K.~Park, ``A case for smartnic-accelerated private
  communication,'' in \emph{4th Asia-Pacific Workshop on Networking}, 2020, pp.
  30--35.

\bibitem{agrawal2012performance}
H.~Agrawal, Y.~Dutta, and S.~Malik, ``Performance analysis of offloading ipsec
  processing to hardware based accelerators,'' in \emph{2012 International
  Symposium on Electronic System Design (ISED)}.\hskip 1em plus 0.5em minus
  0.4em\relax IEEE, 2012, pp. 291--294.

\bibitem{han2010packetshader}
S.~Han, K.~Jang, K.~Park, and S.~Moon, ``Packetshader: a gpu-accelerated
  software router,'' \emph{ACM SIGCOMM Computer Communication Review}, vol.~40,
  no.~4, pp. 195--206, 2010.

\bibitem{li2013gamt}
Y.~Li, D.~Zhang, A.~X. Liu, and J.~Zheng, ``Gamt: a fast and scalable ip lookup
  engine for gpu-based software routers,'' in \emph{Architectures for
  Networking and Communications Systems}.\hskip 1em plus 0.5em minus
  0.4em\relax IEEE, 2013, pp. 1--12.

\bibitem{vasiliadis2014gaspp}
G.~Vasiliadis, L.~Koromilas, M.~Polychronakis, and S.~Ioannidis, ``Gaspp: A
  gpu-accelerated stateful packet processing framework.'' in \emph{USENIX
  Annual Technical Conference}, 2014, pp. 321--332.

\bibitem{wang2023oxdp}
F.~Wang, G.~Zhao, Q.~Zhang, H.~Xu, W.~Yue, and L.~Xie, ``Oxdp: Offloading xdp
  to smartnic for accelerating packet processing,'' in \emph{2022 IEEE 28th
  International Conference on Parallel and Distributed Systems (ICPADS)}.\hskip
  1em plus 0.5em minus 0.4em\relax IEEE, 2023, pp. 754--761.

\bibitem{kim2021linefs}
J.~Kim, I.~Jang, W.~Reda, J.~Im, M.~Canini, D.~Kosti{\'c}, Y.~Kwon, S.~Peter,
  and E.~Witchel, ``Linefs: Efficient smartnic offload of a distributed file
  system with pipeline parallelism,'' in \emph{Proceedings of the ACM SIGOPS
  28th Symposium on Operating Systems Principles}, 2021, pp. 756--771.

\bibitem{kim2018hyperloop}
D.~Kim, A.~Memaripour, A.~Badam, Y.~Zhu, H.~H. Liu, J.~Padhye, S.~Raindel,
  S.~Swanson, V.~Sekar, and S.~Seshan, ``Hyperloop: group-based nic-offloading
  to accelerate replicated transactions in multi-tenant storage systems,'' in
  \emph{Proceedings of the 2018 Conference of the ACM Special Interest Group on
  Data Communication}, 2018, pp. 297--312.

\bibitem{zhang2020fpga}
T.~Zhang, J.~Wang, X.~Cheng, H.~Xu, N.~Yu, G.~Huang, T.~Zhang, D.~He, F.~Li,
  W.~Cao \emph{et~al.}, ``Fpga-accelerated compactions for lsm-based key-value
  store.'' in \emph{FAST}, 2020, pp. 225--237.

\bibitem{Li_Chen_Shen_Wang_Cao_2025}
\BIBentryALTinterwordspacing
F.~Li, Q.~Chen, J.~Shen, X.~Wang, and J.~Cao, ``Performance characteristics and
  guidelines of offloading middleboxes onto bluefield-2 dpu,'' \emph{IEEE
  Transactions on Computers}, vol.~74, no.~2, p. 609–622, Feb. 2025.
  [Online]. Available:
  \url{https://ieeexplore.ieee.org/abstract/document/10756527}
\BIBentrySTDinterwordspacing

\bibitem{wei2023characterizing}
X.~Wei, R.~Cheng, Y.~Yang, R.~Chen, and H.~Chen, ``Characterizing off-path
  $\{$SmartNIC$\}$ for accelerating distributed systems,'' in \emph{17th USENIX
  Symposium on Operating Systems Design and Implementation (OSDI 23)}, 2023,
  pp. 987--1004.

\bibitem{DBLP:journals/corr/abs-2105-06619}
\BIBentryALTinterwordspacing
J.~Liu, C.~Maltzahn, C.~D. Ulmer, and M.~L. Curry, ``Performance
  characteristics of the bluefield-2 smartnic,'' \emph{CoRR}, vol.
  abs/2105.06619, 2021. [Online]. Available:
  \url{https://arxiv.org/abs/2105.06619}
\BIBentrySTDinterwordspacing

\bibitem{Thostrup2022ADE}
\BIBentryALTinterwordspacing
L.~Thostrup, D.~Failing, T.~Ziegler, and C.~Binnig, ``A dbms-centric evaluation
  of bluefield dpus on fast networks,'' in \emph{ADMS@VLDB}, 2022. [Online].
  Available: \url{https://api.semanticscholar.org/CorpusID:252125087}
\BIBentrySTDinterwordspacing

\bibitem{jeong2014mtcp}
E.~Jeong, S.~Wood, M.~Jamshed, H.~Jeong, S.~Ihm, D.~Han, and K.~Park,
  ``$\{$mTCP$\}$: a highly scalable user-level $\{$TCP$\}$ stack for multicore
  systems,'' in \emph{11th USENIX Symposium on Networked Systems Design and
  Implementation (NSDI 14)}, 2014, pp. 489--502.

\bibitem{doca_dma}
``Doca+dma,'' \url{https://docs.nvidia.com/doca/sdk/doca+dma/index.html}.

\bibitem{dpdk}
\BIBentryALTinterwordspacing
{Linux Foundation}, ``Data plane development kit ({DPDK}),'' 2025. [Online].
  Available: \url{https://www.dpdk.org}
\BIBentrySTDinterwordspacing

\bibitem{fstack}
Tencent, ``{F-stack homepage},'' \url{https://www.f-stack.org/}, 2023, [Online;
  accessed 7-Apr-2023].

\bibitem{F_stack_issue}
``F-stack issue,'' \url{https://github.com/F-Stack/f-stack/issues/694/}.

\bibitem{seastar}
``Seastar,'' \url{https://seastar.io/}.

\bibitem{lthread}
``lthread: Multicore / multithread coroutine library.''
  \url{https://lthread.readthedocs.io}.

\bibitem{wrk}
``wrk,'' \url{https://github.com/wg/wrk}.

\bibitem{dpbench}
``dpbench,'' \url{https://github.com/dpbench/dpbench}.

\bibitem{arm_cci400}
\BIBentryALTinterwordspacing
ARM, ``Corelink cci-400 cache coherent interconnect.'' [Online]. Available:
  \url{https://www.arm.com/products/silicon-ip-system/corelink-interconnect/cci-400}
\BIBentrySTDinterwordspacing

\end{thebibliography}





\vfill

\end{document}